\documentclass{article}

\usepackage{amsthm,amsmath,amsfonts,amssymb,mathrsfs}
\usepackage{moreverb}
\usepackage{multirow}
\usepackage{upgreek}
\usepackage[numbers]{natbib} 
\usepackage{graphicx}

\usepackage{geometry}
\usepackage[title]{appendix}

\usepackage[dvips,colorlinks,bookmarksopen,bookmarksnumbered,citecolor=red,urlcolor=red]{hyperref}

\usepackage{bbm}

\newcommand\BibTeX{{\rmfamily B\kern-.05em \textsc{i\kern-.025em b}\kern-.08em
T\kern-.1667em\lower.7ex\hbox{E}\kern-.125emX}}

\def\by{\boldsymbol{y}}
\def\bY{\boldsymbol{Y}}
\def\bx{\boldsymbol{x}}
\def\bX{\boldsymbol{X}}
\def\bu{\boldsymbol{u}}
\def\bz{\boldsymbol{z}}
\def\bZ{\boldsymbol{Z}}
\def\bt{\boldsymbol{t}}
\def\bh{\boldsymbol{h}}

\def\R{\mathbb{R}}
\def\RR{\mathbb{R}^2}
\def\RRR{\mathbb{R}^3}

\def\V{\textrm{Var}}
\def\E{\mathbb{E}}

\def\mB{\mathcal{B}}
\def\mR{\mathcal{R}}

\def\mS{\mathcal{S}}
\def\mK{\mathcal{K}}

\def\bmX{\boldsymbol{\mathcal{X}}}
\def\bmY{\boldsymbol{\mathcal{Y}}}

\def\dint{\,\textrm{d}}

\begin{document}


\title{Tutorial on kernel estimation of continuous spatial and spatiotemporal relative risk with accompanying instruction in R}

\author{Tilman M.~Davies$^{a*}$, Jonathan C. Marshall$^b$, and Martin L. Hazelton$^b$\\ \small
$^{a}$Dept.\ of Mathematics \& Statistics, University of Otago, Dunedin, New Zealand\\
\small $^{b}$Institute of Fundamental Sciences, Massey University, Palmerston North, New Zealand\\
\small $^*$tdavies@maths.otago.ac.nz\normalsize}



\maketitle

\begin{abstract}
	Kernel smoothing is a highly flexible and popular approach for estimation of probability density and intensity functions of continuous spatial data. In this role it also forms an integral part of estimation of functionals such as the density-ratio or ``relative risk'' surface. Originally developed with the epidemiological motivation of examining fluctuations in disease risk based on samples of cases and controls collected over a given geographical region, such functions have also been successfully employed across a diverse range of disciplines where a relative comparison of spatial density functions has been of interest. This versatility has demanded ongoing developments and improvements to the relevant methodology, including use spatially adaptive smoothers; tests of significantly elevated risk based on asymptotic theory; extension to the spatiotemporal domain; and novel computational methods for their evaluation. In this tutorial paper we review the current methodology, including the most recent developments in estimation, computation and inference. All techniques are implemented in the new software package \texttt{sparr}, publicly available for the \texttt{R} language, and we illustrate its use with a pair of epidemiological examples.
\end{abstract}

\section{Introduction}
Kernel smoothing is a well-established nonparametric approach to estimation of continuous density functions based on a sampled dataset. It can cope in a flexible way with inhomogeneous dispersions of observations that are not readily identified as draws from analytically tractable parametric families of distribution. This has rendered kernel density estimation particularly convenient for applications in spatial and spatiotemporal statistics, where high heterogeneity of observed data is commonplace.


While estimation of standalone spatial and spatiotemporal density (or \emph{intensity}---yielding the expected number of points per unit area) is in and of itself an important end goal, it is important to note that such an estimate also forms a key ingredient in the subsequent estimation of more complicated descriptors, functionals, and models of the process(es) at hand. One particularly useful technique is the kernel \emph{density-ratio} or \emph{relative risk} estimator, originally proposed by Bithell \cite{bit:1990,bit:1991} as an exploratory tool to investigate the fluctuation in disease incidence relative to a spatially heterogeneous at-risk population distribution. Since being developed further with epidemiological motivations by Kelsall \& Diggle \cite{keldig:1995a,keldig:1995b}, the relative risk estimator has been successfully deployed to answer research questions in a wide range of disciplines. Some recent examples taken from the varied literature include Bakian et al. \cite{baketal:2015}, who studied the relative risk of autistic disorders in Utah; Campos \& Fedigan \cite{camfed:2014} studied different alarm calls of capuchin monkeys in Costa Rica; Smith et al. \cite{smietal:2015} and Bevan \cite{bev:2012} used the kernel relative risk estimator to examine various patterns of archaeological finds; and microscopic muscle fibre distributions were compared in \cite{davetal:2013}.

This versatility has in turn prompted methodological developments. A novel point-wise test of statistically significant risk fluctuations without the need for simulation was suggested in \cite{hazdav:2009}. Utilisation of spatially adaptive smoothing, which allows the bandwidth of each kernel to vary depending on the position of a given point, followed in \cite{davhaz:2010} with related work in \cite{davetal:2016}. The spatial relative risk function was generalised in \cite{zhaetal:2011} and \cite{ferhaz:2014}, providing strategies for smoothing continuous spatiotemporal data, estimation of relative risk in space-time, and conducting related significance tests.


Alongside this growing suite of methods are ever-present issues in nonparametric smoothing, such as bandwidth selection and edge-correction, as well as computational complexities of implementation.  The overarching objective of this work is therefore to provide an instructional overview of the current state-of-the-science of the kernel density estimator of relative risk in space and space-time. Focus is on the practical aspects of the methodology, and we dovetail the theory with demonstrations of its implementation via our newly developed software package publicly available for the \verb|R| language.

\subsection{Motivating Examples}\label{sec:mot}
We introduce a pair of motivating examples in epidemiology to illustrate the types of problems that benefit from this methodology.

\begin{figure}[hbpt]
\centering
\includegraphics[width=0.5\textwidth]{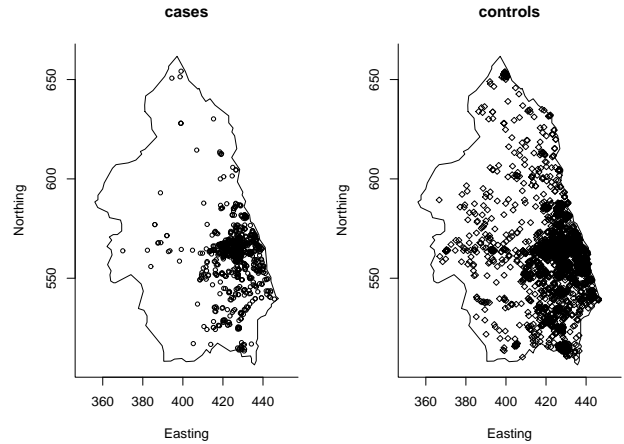}
\vspace{-4mm}
\caption{PBC cases and sampled at-risk controls in a geographical region of northeast England.}\label{fig:pbcdata}
\end{figure}

The first is based on the spatial distribution of 761 cases of primary biliary cirrhosis (PBC) recorded in an area of northeast England formed by several adjacent health regions. Along with 3020 controls representing the at-risk population, these data were first presented and analysed by Prince et al. \cite{prinetal:2001}. The case and control data are plotted in Figure \ref{fig:pbcdata}. Is any local aggregation of cases merely a reflection of the underlying at-risk population dispersion? If not, where and how does the distribution of PBC differ significantly from the background population?

\begin{figure}[hbpt]
	\centering
	\includegraphics[width=1\textwidth]{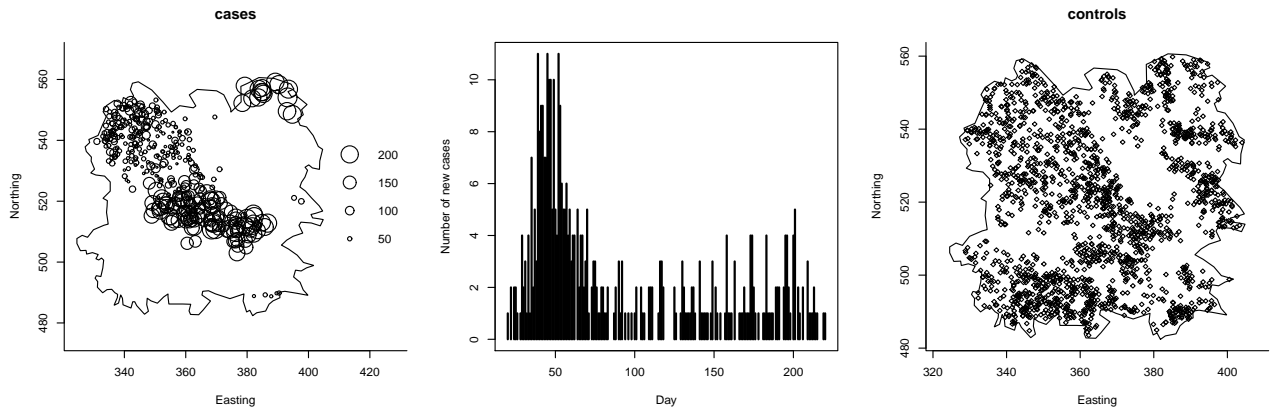}
	\vspace{-4mm}
	\caption{FMD cases, as a spatial and temporal plot on the left and middle, and sampled at-risk controls on the right, in Cumbria, England, during the 2001 outbreak. The size of the plotted cases is proportional to the day of infection. Data have been jittered and randomly thinned to preserve anonymity.}\label{fig:fmddata}
\end{figure}

The second example concerns the spread of 410 farms affected by foot-and-mouth disease (FMD) over space and time during the 2001 outbreak in Cumbria, England (see \cite{keeeatal:2001,lawzho:2005}). The sampled locations of 1866 uninfected farms are also recorded, and the data are shown in Figure \ref{fig:fmddata} (for reasons of confidentiality, the observations have been jittered and randomly thinned by an unspecified amount). How does the spatial risk of infection change over the duration of the study period? Are we able to detect emerging ``hotspots'' of disease as they occur?

\subsection{\texttt{R} Package ``\texttt{sparr}''}
Alongside the review of the statistical techniques, we introduce the newly redeveloped software package \verb|sparr| (\textit{spa}tial \textit{r}elative \textit{r}isk), which incorporates all key methods discussed herein and aims to provide researchers with the tools necessary to answer research questions such as those posed in the motivating examples. The package is freely available for the \verb|R| language \cite{rcite} on the Comprehensive \verb|R| Archive Network (CRAN) at \small\url{https://cran.r-project.org}\normalsize, and the user can install the library directly from an \verb|R| prompt with a call to

\small
\begin{verbatim}
R> install.packages("sparr")
\end{verbatim}
\normalsize
followed by loading with

\small
\begin{verbatim}
R> library("sparr")
\end{verbatim}
\normalsize
to gain access to its functionality.
 
Originally released in 2010, the \verb|sparr| package has to date possessed relatively limited functionality dealing with purely spatial analyses, which were discussed in the vignette \cite{davetal:2011}. The new \verb|sparr| (versions $<$ 2.1-10) has been completely redesigned with notable extensions as follows.
\begin{itemize}
\item \textbf{Spatial and spatiotemporal estimation:} \verb|sparr| is equipped to deal with fixed- as well as adaptive-bandwidth kernel estimation of spatial (2D) density and relative risk functions based on data observed on an irregularly shaped study region; including the incorporation of edge correction factors. Functionality is also present for kernel density and relative risk estimation of spatiotemporal (3D) data and appropriate edge correction.
\item \textbf{Bandwidth selection:} A number of relevant bandwidth selectors have been made available in \verb|sparr|. These range from simple rule-of-thumb selectors we have previously found useful in these endeavours, to spatial and spatiotemporal adaptations of classical leave-one-out cross-validation techniques and computationally intensive bootstrap methods; many of which must be corrected for edge-effects on the relevant domain. These methods incorporate the initial pursuits made thus far to address the difficult problem of bandwidth selection for spatially adaptive and spatiotemporal bandwidth selection, as well as jointly optimal bandwidth selection for relative risk function estimation. Noteworthy is the fact this remains an active area of research.
\item \textbf{Multi-scale adaptive smoothing:} Recent advancements to evaluation of the spatially adaptive kernel estimator \cite{davbad:2017} have been made available in \verb|sparr|, in part to assist the expensive operations associated with repeated evaluation of such density estimates by producing so-called \emph{multi-scale} estimates that simultaneously smooth relevant estimates at different bandwidths. 
\item \textbf{Speed improvements and \texttt{spatstat} compatibility:} A universal motivation has been to ease access to these computationally expensive techniques, and we have aimed to minimise computation time for all functions. Additionally, the usage and output of relevant functions has been designed to enhance their compatibility with \verb|spatstat| \cite{badturn:2005,badetal:2015}, a comprehensive \verb|R| package for spatial point pattern analysis; mainly through use of the \verb|ppp|, \verb|owin|, and \verb|im| object classes thereof.
\end{itemize}

As part of this paper we provide practical code demonstrations detailing \verb|sparr|'s use. The complete code used to produce all Figures and estimates in this work is supplied as supplementary material in a downloadable \verb|R| script file.

\subsection{Article Structure}
The tutorial is structured as follows. In Section \ref{sec:kde} we review the kernel density estimator for spatial and spatiotemporal data, including the concept of spatially adaptive smoothing and the necessity of edge-correction factors. Bandwidth selection strategies for these density estimators with respect to the type of applications of interest are discussed in Section \ref{sec:banddens}. The variants of the spatial relative risk function, aspects of jointly optimal bandwidth selection, and associated significance tests for constructing tolerance contours are detailed in Section \ref{sec:srr}, followed by a code break with some examples in Section \ref{sec:srrcode}. In Section \ref{sec:strr} we detail estimation and testing of spatiotemporal relative risk, and a corresponding set of coded examples appear in Section \ref{sec:strrcode}. Brief details of topics related to the computation of adaptive kernel estimates and additional visualisation techniques are given in Section \ref{sec:adv}, and concluding remarks on future research appear in Section \ref{sec:conc}. Appendix \ref{app:sparr} contains an index of the main functions of \verb|sparr|, linking them to the methodology and equations in the paper.

\section{Kernel Density Estimation}\label{sec:kde}
Suppose we observe $n$ points in a 2-dimensional (2D) space; $\bX=\{\bx_1,\ldots,\bx_n\}$. It is always the case in practice that these points are observed in a bounded subset $W$ of the plane; $\bX\in W\subset\RR$, and we refer to $W$ as the spatial \emph{window} or \emph{study region}. This boundedness is an important feature of the data, and affects all stages of estimation and inference.

Assuming $\bX$ arises from a probability density function $f$ with domain $W$, the goal of kernel density estimation is to estimate $f$, the origin of the data.

\subsection{Spatial: Fixed Bandwidth}\label{sec:kdefix}
The classical fixed-bandwidth kernel estimate of $f$ given $\bX$ is written as
\begin{equation}\label{eq:kdefix}
\tilde{f}_h(\by|\bX) = n^{-1}h^{-2}\sum_{i=1}^{n} K\left(\frac{\by-\bx_i}{h}\right)q_h(\by|W)^{-1};\qquad\by\in W,
\end{equation}
where $h>0$ is the scalar \emph{bandwidth} or \emph{smoothing parameter}; $K(\cdot)$ is a 2D, zero-centered, radially symmetric probability density referred to as the \emph{kernel function}; and $q_h(\cdot)$ is an \emph{edge-correction} factor we shall discuss in greater detail momentarily. The kernel estimator works by centering a kernel `bump' of probability weight $1/n$ atop each observation, and summing the resulting contributions at each evaluation coordinate $\by\in W$. 

Arguably the most important component of (\ref{eq:kdefix}) is the bandwidth $h$. This directly controls the spread of each kernel---while the contributory probability remains $1/n$, a large $h$ spreads the kernel wider and lower; and a small $h$ focusses the weight so that each kernel becomes narrower and taller. Consequently, the overall smoothness of the density estimate $\tilde{f}_h$ is affected, and thus \emph{bandwidth selection}, the aim of which is to strike an optimal balance between under- and over-smoothing given $\bX$, is important. Our definition in the current setting is to supply $h$ as a scalar---which provides \emph{isotropic} smoothing---such that the kernel is smoothed by the same amount parallel to both horizontal and vertical axes. More generally, the bivariate kernel density estimator sees $h$ expressed as a $2\times 2$ bandwidth matrix $H$, which permits both the axis-specific smoothing as well as the relative orientation of the kernel itself to be altered (see e.g.\ \cite{wj:1993,wj:1995}). For the spatial applications we consider, however, isotropic smoothing is typically sufficient.

Choice of the kernel function itself is generally acknowledged to be of secondary importance to the bandwidth; see \cite{wj:1995}. It therefore tends to be set with practical reasons in mind, and one of the most popular choices is the Gaussian kernel---which is what we shall use herein. Coupled with convenient theoretical properties, the infinite tails of the Gaussian kernel can for example help with smoothing highly heterogeneous point patterns where there exist sizeable areas of $W$ devoid of observations.

The final aspect of the 2D kernel estimator above concerns edge-correction. Observations of $\bX$ that lie near the boundary of $W$ inevitably lose some measurable probability weight over the edge; leading to a negative bias in the final estimate. This \emph{boundary bias} has been shown to be severe enough to warrant correction based on the asymptotic properties of the estimator \cite{dig:1985,jon:1993,marhaz:2010}. For planar point patterns, the simplest way to achieve this is a rescaling of the estimate at hand. The correction factor shown in (\ref{eq:kdefix}) is defined as
\begin{equation}\label{eq:qfix}
q_h(\by|W)=h^{-2}\int_W K\left(\frac{\bu-\by}{h}\right)\dint\bu;\qquad\by\in W
\end{equation}
which can be interpreted as the proportion of the kernel weight that falls within $W$ for a kernel centered at $\by$ with bandwidth $h$. A subtly different yet important alternative correction factor is obtained by rescaling at each observation instead of evaluation coordinate i.e.\ by replacing $q_h(\by|W)$ in (\ref{eq:kdefix}) by $q_h(\bx_i|W)$. The version based on $\by$ can often be convenient for computational reasons because it can be evaluated independently of the main kernel sum; though where possible the version based on $\bx_i$ is preferred for numerical precision. Hereinafter, we take cues from the literature and refer to correction based on evaluation coordinate $\by$ as ``uniform'' edge-correction; and correction based on observation $\bx_i$ as ``Diggle'' edge-correction. For further details on this aspect of smoothing spatial point patterns, see \cite{dig:1985,jon:1993} as well as \cite{dig:2010} and references therein.

\begin{figure}[hbtp]
\centering
\includegraphics[width=0.6\textwidth]{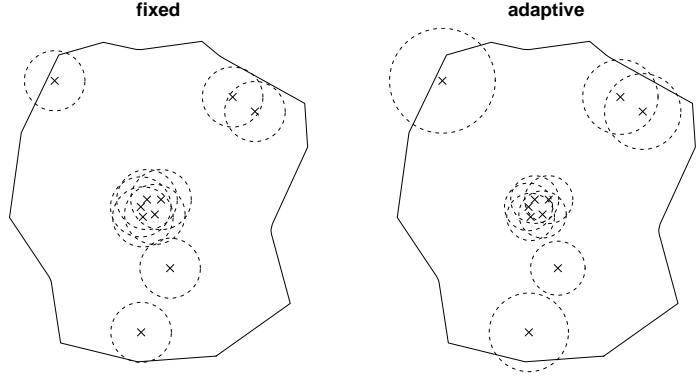}
\caption{Comparison of fixed and adaptive kernel spreads for respective density estimates given 10 hypothetical observations and an associated study window.}\label{fig:kde}
\end{figure}

For illustration of the above ideas, Figure \ref{fig:kde} provides a toy example of 10 planar observations in a hypothetical study region. Superimposed discs delineate the isotropic kernel at one bandwidth (in the case of the Gaussian kernel, this is also equivalent to one standard deviation) away from each owning point. Visible is the equality of the spread of each of the 10 kernels for the fixed-bandwidth estimator, as well as the potential for substantial probability weight to be lost over the boundary for some observations.

\vspace{-5mm}

\subsection{Spatial: Adaptive Bandwidth}\label{sec:kdeada}
By relaxing the requirement that the bandwidth of each of the $n$ kernels be constant, we obtain a \emph{sample-point adaptive} estimator, certain forms of which have been shown to possess both theoretical and practical advantages over their fixed-bandwidth counterpart. The adaptive estimator takes the form
\begin{equation}\label{eq:kdeada}
\hat{f}_{h_0}(\by|\bX) = n^{-1}\sum_{i=1}^{n}h(\bx_i;f)^{-2} K\left(\frac{\by-\bx_i}{h(\bx_i;f)}\right)q_{h(\by)}(\by|W)^{-1};\qquad\by\in W;
\end{equation}
the key difference from (\ref{eq:kdefix}) being definition of the smoothing bandwidth as a function of a coordinate. The question arises as to how one might best ``adapt'' the bandwidths via $h(\cdot)$.

The preeminent approach for sample-point adaptation is due to Abramson \cite{abram:1982}. Based on the work in \cite{brietal:1977}, Abramson suggested setting $h(\bu;f)=h_0f(\bu)^{-1/2}$; that is, bandwidths being inversely proportional to the square root of the target density itself. In practice, in line with Silverman \cite{sil:1986}, Abramson's smoothing regimen is implemented as
\begin{equation}\label{eq:hu}
h(\bu;f)=h_0\tilde{f}_{\tilde{h}}(\bu|\bX)^{-\frac{1}{2}}\gamma_f^{-1},
\end{equation}
where $\tilde{f}$ is a pilot estimate of the unknown density constructed via (\ref{eq:kdefix}) with a fixed \emph{pilot bandwidth} $\tilde h$; $h_0$ is an overall smoothing parameter for the variable bandwidths referred to as the \emph{global bandwidth}; and $\gamma_f=\exp\left\{n^{-1}\sum_i\log\left(\tilde{f}_{\tilde{h}}(\bx_i|\bX)^{-1/2}\right)\right\}$ is the geometric mean of the inverse-density bandwidth factors (in place so $h_0$ can be considered on the same scale as any fixed bandwidth for the same data; see \cite{sil:1986}).

From (\ref{eq:hu}), it is clear how the variable bandwidths behave. In areas of high point density (where $\tilde{f}$ is large), the resulting bandwidth will be small, while in areas of low point density (where $\tilde{f}$ is small) the resulting bandwidth will be large. We are thus able to reduce smoothing in spatial sub-regions of $W$ where there are many observations in order to capture greater detail in the resulting density estimate, whilst increasing smoothing in other areas where a relative lack of observations heightens uncertainty. The adaptation imposed by (\ref{eq:hu}) has been shown under certain conditions to yield improvements to asymptotic bias when compared to the fixed-bandwidth estimator \cite{abram:1982,halmar:1988}.

To ensure the pilot density $f$ is sufficiently bounded away from zero so as to prevent variable bandwidths becoming excessively large \cite{halmar:1988}, in practice we replace $\tilde f$ with $\min\left\{\tilde{f}_{\tilde{h}}(\cdot)^{-\frac{1}{2}},\tau\gamma_f\right\}$, where we find a value of $\tau \approx 5$ suits in many examples.


Bandwidth selection in the adaptive setting therefore targets the global bandwidth $h_0$; this is discussed alongside fixed bandwidth selection in Section \ref{sec:banddens}.

Edge correction for the adaptive estimator may be carried out in much the same way as outlined earlier; for theoretical details see \cite{marhaz:2010}. The uniform factor in (\ref{eq:kdeada}) is 
\begin{equation}\label{eq:qada}
q_{h(\by;f)}(\by|W)=h(\by;f)^{-2}\int_W K\left(\frac{\bu-\by}{h(\by;f)}\right)\dint\bu;\qquad\by\in W, 
\end{equation}
with the alternative Diggle factors given accordingly as $q_{h(\bx_i;f)}(\bx_i|W)$ inside (\ref{eq:kdeada}).

As a visual comparison to fixed bandwidth estimation, the right image of Figure \ref{fig:kde} shows the kernel spread at one bandwidth away from each of the 10 hypothetical observations for the adaptive estimator. Note the effect of Abramson's smoothing regimen---kernel smoothing reduced in pockets of relatively high point density, and increased for those more isolated observations.

\subsection{Spatiotemporal}\label{sec:st}
Suppose in addition to observing the spatial locations $\bx_i$ we are also availed the time of occurrence of each event, $t_i$. The density $f$ that we now seek to estimate is \emph{trivariate}---we are adding an extra dimension to the problem when compared to the spatial-only setting---requiring simultaneous smoothing in space and time.

Define our data set as  $\bmX=\{(\bx_1,t_1),\ldots,(\bx_n,t_n)\}$; $i=1,\ldots,n$. It is assumed the spatial margin is contained by $W\subset\RR$ as earlier, with timestamps falling in some well-defined interval $T\subset\R$. We use the estimator detailed in \cite{ferhaz:2014}, given as
\begin{equation}\label{eq:kdest}
\breve{f}_{h,\lambda}(\bz,s|\bmX) = n^{-1}h^{-2}\lambda^{-1}\sum_{i=1}^{n} K\bigg(\frac{\bz-\bx_i}{h}\bigg)L\bigg(\frac{s-t_i}{\lambda}\bigg)q_h(\bz|W)^{-1}w_\lambda(s|T)^{-1};\qquad\bz\in W,\quad s\in T,
\end{equation}
where $K(\cdot)$, $h$, and $q_h(\cdot|W)$ are the 2D kernel, isotropic bandwidth, and edge-correction factor for the spatial margin defined exactly as in Section \ref{sec:kdefix}. The temporal margin is smoothed via the univariate kernel $L$ (taken to be a zero-centered symmetric pdf), bandwidth $\lambda>0$, and is corrected for edge effects (imposed by the temporal bounds of $T$) by
\begin{equation}\label{eq:wt}
w_\lambda(s|T) = \lambda^{-1}\int_T L\left(\frac{t-s}{\lambda}\right)\dint t
\end{equation}
i.e.\ the proportion of probability weight of $L$ centered on $s$ with bandwidth $\lambda$ that lies within $T$. Thus, implementation of the spatiotemporal smoother (\ref{eq:kdest}) is subject to the same considerations as the spatial-only estimator (\ref{eq:kdefix}), with choice of bandwidth $(h,\lambda)$ of primary interest. 

\begin{figure}[hbpt]
\centering
\includegraphics[width=0.23\textwidth]{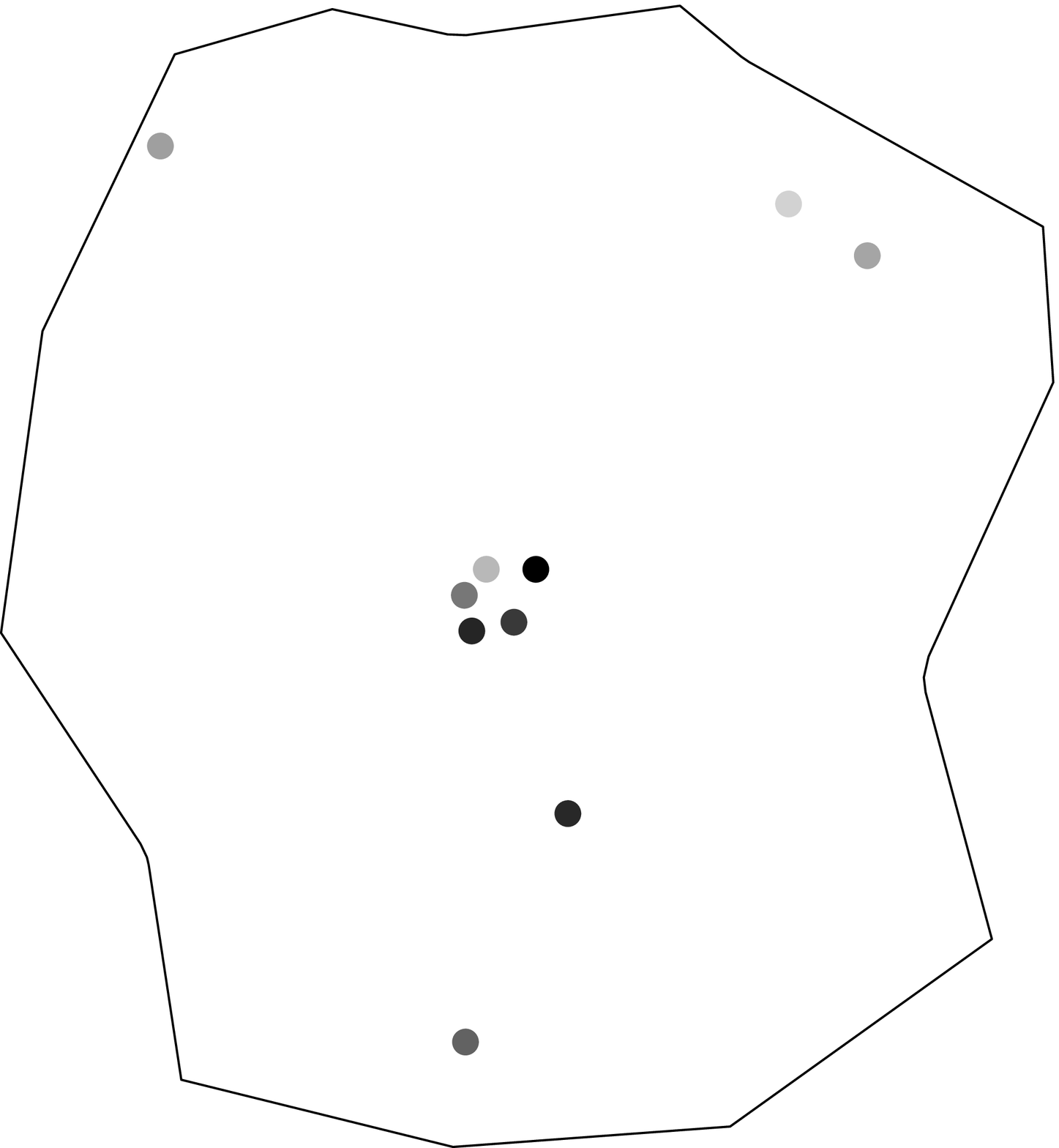}
\hspace{10mm}
\includegraphics[width=0.25\textwidth]{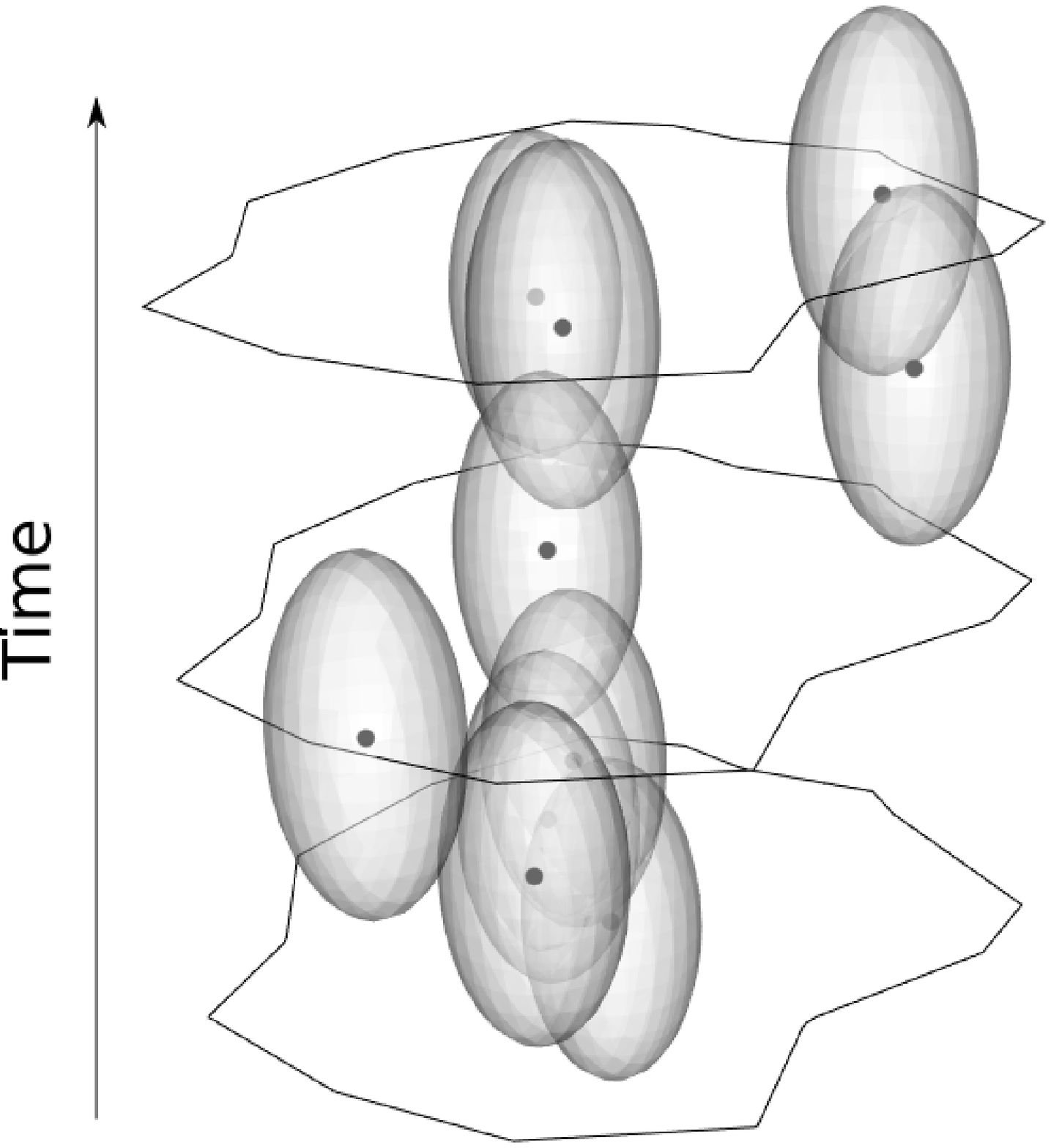}
\caption{Plots of 10 hypothetical spatiotemporal observations (left, where dark $\rightarrow$ light denotes increasing time); and 3D kernel spreads in space and time of each point (right, drawn at a one particular level as translucent grey isosurfaces). In the latter image the spatial region $W$ is superimposed at the bounds and midpoint of the temporal interval $T$, where an increase in time is interpreted as moving up along the vertical axis.}\label{fig:toyst}
\end{figure}

The idea of spatiotemporal smoothing is illustrated in Figure \ref{fig:toyst}, using the same $10$ hypothetical spatial points and study window shown in Figure \ref{fig:kde}. Here, we assume each point is marked with a particular time, shown on the left. Smoothing as per (\ref{eq:kdest}) is performed via a trivariate kernel (the result of the product of $K$ and $L$). The 3D kernels, which in our implementation are set as the 2D Gaussian for $K$ and 1D Gaussian for $L$, are drawn at a particular numeric level the same distance away from each owning data point. The planar window $W$ is drawn at the bounds of the temporal margin $T$, which is plotted along the vertical axis, as well as at the midpoint of $T$ to emphasise the unchanging nature of the spatial region over time. The need to edge-correct in both the spatial and temporal margins throughout the domain $W\times T$ via $q_h(\cdot)$ and $w_\lambda(\cdot)$ is clear to see.

\vspace{-5mm}

A final issue unique to the spatiotemporal setting is the concept of the time-dependent \emph{conditional} spatial density over $W$. In its current form, (\ref{eq:kdest}) provides the \emph{joint} density estimate over space and time, in the sense that
\[
\int_W\int_T \breve{f}_{h,\lambda}(\bz,t|\bmX) \dint t \dint \bz = 1.
\]
If the interest is an inspection of the spatial density estimate for a given time $t$, it can be more convenient to condition on the latter, normalising the surface such that the temporal slice itself integrates to $1$ over $W$. Specifically, we write
\begin{equation}\label{eq:kdestcond}
\breve{f}_{h,\lambda}(\bz|s=t,\bmX)=\frac{\breve{f}_{h,\lambda}(\bz,t|\bmX)}{\bar{f}_\lambda(t|\bt)};\qquad t\in T,\quad\by\in W,
\end{equation}
where $\bt=\{t_1,\ldots,t_n\}$ are the timestamps of the $n$ observations, and
\begin{equation}\label{eq:ft}
\bar{f}_\lambda(t|\bt)=n^{-1}\lambda^{-1}\sum_{i=1}^{n}L\left(\frac{t-t_i}{\lambda}\right)w_\lambda(t|T)^{-1}
\end{equation}
is the (edge-corrected) univariate kernel density estimate of the temporal margin of the data $\bmX$. Through (\ref{eq:kdestcond}) we therefore have
\[
\int_W \breve{f}_{h,\lambda}(\by|s=t,\bmX) \dint \by = 1;\qquad t\in T.
\]

The conditional density can be particularly useful in estimates of spatiotemporal relative risk, where the changing nature of the relative density over time is under scrutiny, as well as in more general data exploration pursuits. We revisit this topic in Section \ref{sec:strr}.

\section{Bandwidth Selection}\label{sec:banddens}
Bandwidth selection methods, all of which aim to provide some `optimal' level of smoothing given the data at hand, have long been a key research focus in the literature. A host of different options have been discussed; a number being described in texts and review papers, see for example \cite{sil:1986,caoetal:1994,wj:1995,jonetal:1996} and references therein.

The lion's share of methods have targeted univariate density estimation problems. A number of complications present themselves when considering bandwidth selection for spatial and spatiotemporal data in the context of the current work. Theoretically, the issue is one of bandwidth selection for the multivariate kernel estimator, for which a focus on bandwidth matrix parameterisation (see \cite{wj:1993} for example) and methods tailored to selection of their components by extending univariate approaches have been successful (\cite{saietal:1994,duohaz:2005}). 

That said, relatively little attention has been paid to important practical aspects of the density estimation at hand, a notable issue being that of edge-correction. The corrective factors alter the asymptotic properties of the kernel estimator upon which many existing selectors are based, and thus care needs to be taken in directly applying such methods for estimation of spatial and spatiotemporal densities. Furthermore, the unique complexities of the adaptive estimator, as well as the difficult issue of choosing `jointly optimal' bandwidths for estimation of spatial and spatiotemporal relative risk surfaces, mean the development of appropriate bandwidth selectors remain an important active area of research.

While a full review of selection methods is beyond the scope of this work, we present here details of some techniques we have found to be useful for the type of density estimation of interest.

\subsection{Rules-of-Thumb}\label{sec:hthumb}
Simple rule-of-thumb methods offer fast, rough indications of what appropriate bandwidths might look like. They are crude in the sense that any data-driven quantification of the distribution of interest is typically limited to summary measures of spread.

The \emph{normal scale rule} is one such example. Under the assumption that the true density of interest $f$ is a bivariate normal, then it can be shown for the fixed-bandwidth kernel estimator using a Gaussian kernel that
\begin{equation}\label{eq:hns}
h_{\textrm{NS}}=\hat{\sigma}n^{-\frac{1}{6}}
\end{equation}
is the asymptotically optimal bandwidth to use, where $\hat{\sigma}$ is a scalar estimate of spread, such as the mean of the axis-specific standard deviations of the observations. To guard against adverse effects of potential outliers, we follow Silverman \cite{sil:1986} and use 
\begin{equation}\label{eq:sighat}
\hat{\sigma}=\min\left\{\frac{s_1+s_2}{2},\frac{\textrm{IQR}_1/1.34+\textrm{IQR}_2/1.34}{2}\right\},
\end{equation}
where $s_1$, $s_2$ and IQR$_1$, IQR$_2$ are the axis-specific standard deviations and interquartile ranges respectively. While $h_\textrm{NS}$ can be expected to perform well if our data are at least approximately normal, this is rarely the case in practice, and so general application of (\ref{eq:hns}) is usually limited to basic exploration of the data.

An alternative scale-based measure is that of the \textit{maximal smoothing principle} of Terrell \cite{ter:1990}. As its title suggests, this rule is based on a relatively straightforward upper bound on the optimal bandwidth with respect to the asymptotic mean integrated squared error of the density estimate, and provides an \emph{oversmoothing} bandwidth. For a fixed bandwidth density estimate of dimension $d$ based on a sample of size $n$, we have 
\begin{equation}\label{eq:hosd}
h_{\textrm{OS}(d)}=\hat{\sigma}\left\{\frac{(d+8)^{(d+6)/2}\pi^{d/2}R(K)}{16n\Gamma[(d+8)/2](d+2)}\right\}^{1/(d+4)}
\end{equation}
where $R(K)=\int K^2$, $\Gamma[\,\cdot\,]$ is the gamma function, and $\hat{\sigma}$ is a scalar measure of spread. For a spatial density estimate using the Gaussian kernel, (\ref{eq:hosd}) simplifies to
\begin{equation}\label{eq:hos}
h_\textrm{OS}=h_{\textrm{OS}(2)}=\hat{\sigma}\left(\frac{625}{384n}\right)^{1/6},
\end{equation}
with it again being possible to set $\hat{\sigma}$ as per (\ref{eq:sighat}). For general density estimation tasks, we would expect $h_{\textrm{OS}}$ to provide bandwidths that are too smooth; however, this behaviour can to a certain extent be desirable for kernel estimation of relative risk functions. We clarify this remark in Section \ref{sec:srr}.

Use of fixed bandwidth scale rules such as $h_\textrm{NS}$ and $h_\textrm{OS}$ can, as rough guesses, be used to set $h_0$ in an adaptive kernel estimate as per (\ref{eq:kdeada}); in recent work it has been shown that specific expressions analagous to (\ref{eq:hns}) for the adaptive estimator are not possible for theoretical reasons \cite{davetal:2017}. Furthermore, combining a 2D and 1D version of either of the above, we may also use the resulting bandwidths as $h$ and $\lambda$ in a spatiotemporal density estimate given by (\ref{eq:kdest}). The cautionary notes regarding the approximate and exploratory nature of such bandwidths and the densities that result remain an important consideration in these settings.

\subsection{Least Squares Cross-Validation}\label{sec:lscv}
A very common overall measure of accuracy of a kernel estimate with respect to its target density is that of the \emph{mean integrated squared error} (MISE), which for the fixed bandwidth estimator is given as
\begin{equation}\label{eq:mise}
\textrm{MISE}(\tilde{f}_h) =\E\left[\int_W \left\{\tilde{f}_h(\by|\bX)-f(\by)\right\}^2\dint\by\right].
\end{equation}
Density estimates that lie `closer' to the true $f$ over its domain provide smaller MISEs. In theory, it is therefore sensible to use the MISE as a criterion for selection of an appropriate $h$.

\emph{Least squares cross-validation} (LSCV) does just that; based on an unbiased estimate of (\ref{eq:mise}) modulo an irrelevant additive constant. We find
\[
h_\textrm{LSCV}=\textrm{argmin}_h\left[\textrm{LSCV}(h|\bX)\right]
\]
where
\begin{equation}\label{eq:lscv}
\textrm{LSCV}(h|\bX)=\int_W \tilde{f}_h(\by|\bX)^2\dint\by - 2n^{-1}\sum_{i=1}^{n}\tilde{f}_h(\bx_i|\bX_{[-i]})
\end{equation}
and the latter term above is a leave-one-out estimate evaluated at the $i$th data point location following its omission from the data set, denoted by $\bX_{[-i]}$.

Dealing with direct density estimates of the data, the LSCV criterion readily incorporates any required edge-correction factors. Furthermore, it can be applied with little modification to the spatially adaptive and spatiotemporal estimators. By replacing $\tilde{f}_h$ in (\ref{eq:lscv}) with $\hat{f}_{h_0}$, optimisation can be performed with respect to the global bandwidth $h_0$. In the spatiotemporal case, we replace $\tilde{f}_h$ with $\breve{f}_{h,\lambda}$, and optimisation is 2-dimensional over both $h$ and $\lambda$.

The theoretical appeal of LSCV aside, in practice the performance of this bandwidth selector can on occasion be disappointing, particularly so for spatial data. We have found a tendency for LSCV to select rather small bandwidths in all fixed, adaptive, and spatiotemporal scenarios, and the variability associated with the leave-one-out operation can cause numerical problems when applied to highly heterogeneous data sets. Nevertheless, it is an important classical bandwidth selection method that at the very least provides a benchmark for comparison with other approaches. For further details on LSCV, see Section 3.3 of Wand \& Jones \cite{wj:1995} and references therein.

\subsection{Likelihood Cross-Validation}\label{sec:hlik}
A different optimisation criterion is gained by thinking about the estimated density itself as a likelihood function, and maximising its logarithm with respect to the bandwidth by taking a leave-one-out average; see Section 3.4.4 of Silverman \cite{sil:1986}. Operationally, it is similar to LSCV. The result is given by
\[
h_\textrm{LIK}=\textrm{argmax}_h\left[\textrm{LIK}(h|\bX)\right]
\]
where
\begin{equation}\label{eq:lik}
\textrm{LIK}(h|\bX)=n^{-1}\sum_{i=1}^n\log\left[\tilde{f}_h(\bx_i|\bX_{[-i]})\right].
\end{equation}
Like LSCV, the LIK selector incorporates edge-correction and can be applied directly for selection of $h_0$ in adaptive smoothing and selection of $h$ and $\lambda$ in spatiotemporal smoothing. That said, it also suffers from similar practical drawbacks and can be sensitive to the effects of outlying observations.

\subsection{Bootstrapping}\label{sec:hboot}
The practical problems noticed with leave-one-out selectors like LSCV and LIK led to the development of methods that could to a certain extent curb the high variability associated with those approaches. The idea of using \textit{bootstrap} resampling to choose a smoothing bandwidth, first suggested in the univariate setting by Taylor \cite{tay:1989} (see also \cite{farjhu:1990}), is one such method that falls under an umbrella of concepts termed \emph{smoothed cross-validation} by Hall et al. \cite{haletal:1992}.

The bootstrap approach is most commonly used in conjunction with the MISE (\ref{eq:mise}) as a target criterion. The basic idea is to construct a `reference' density estimate of the data at hand, repeatedly simulate data from that reference density, and calculate the empirical integrated squared error at each iteration; doing so at different bandwidths. The bandwidth that minimises the bootstrap-estimated MISE is taken as the optimal value. For the fixed bandwidth estimator, we write the optimal bandwidth as
\[
h_\textrm{BOOT}=\textrm{argmin}_h\left[\textrm{BOOT}(h|\bX,\eta)\right]
\]
where $\eta$ is the \emph{reference bandwidth} and
\begin{equation}\label{eq:boot}
\textrm{BOOT}(h|\bX,\eta)=\mathbb{E}_\star\left[\int_W \left\{\tilde{f}_h(\by|\bX^\star)-\tilde{f}_\eta(\by|\bX)\right\}^2\dint\by\right].
\end{equation}
Here, $\bX^\star\sim\tilde{f}_\eta(\by|\bX)$, and $\mathbb{E}_\star$ denotes expectation with respect to the generated data $\bX^\star$.

When the Gaussian kernel function $K$ is used in both reference and generated density estimates, Taylor \cite{tay:1989} cleverly shows that (\ref{eq:boot}) can be evaluated analytically and does not actually require generating samples $\bX^\star$ from the reference density, which leads to tremendous computational savings. In our currently ongoing research pursuits, we have shown that these results can only be partially reproduced in the presence of edge-correction factors for fixed bandwidth spatial and spatiotemporal density estimates, but this nevertheless still improves over `manual' resampling in terms of computational burden. In the same way as the LSCV and LIK selectors, application of (\ref{eq:boot}) to the spatiotemporal domain involves optimisation over both $h$ and $\lambda$; in this case given a pair of reference bandwidths $\eta$ and $\nu$ respectively.

Analytic simplification of (\ref{eq:boot}) is not generally accessible for choice of the global smoothing parameter $h_0$ for a spatially adaptive kernel density estimate. In this instance we must revert to manual evaluation of the empirical MISE, which is greatly eased by recent advancements to the computation of the adaptive kernel estimator in Davies \& Baddeley \cite{davbad:2017} by way of multiscale estimation (briefly discussed in Section \ref{sec:multi}). Specifically, in the adaptive case we have 
\begin{equation}\label{eq:bootada}
\textrm{BOOT}(h_0|\bX,\eta) =J^{-1}\sum_{j=1}^{J}\left[\int_W \left\{\hat{f}_{h_0}(\by|\bX^\star_j)-\hat{f}_\eta(\by|\bX)\right\}^2\dint\by\right],
\end{equation}
where a total of $J$ bootstrap samples $\bX^\star_1,\ldots,\bX^\star_J$ are generated from the reference density $\hat{f}_\eta$ for each trialled value of $h_0$.

A natural question is how one should choose the reference bandwidth $\eta$ (and $\nu$ in the spatiotemporal case). Taylor \cite{tay:1989} simply sets this to be equal to the target bandwidth, allowing it to vary in the minimisation operations. Faraway and Jhun \cite{farjhu:1990} use the LSCV bandwidth obtained via (\ref{eq:lscv}). In our implementations, however, we take note of comments in \cite{haletal:1992} which recommend a generous level of smoothing be used for the reference density. Use of the oversmoothing bandwidth in (\ref{eq:hosd}) is one potential strategy.

\subsection{Other Methods}
There are a number of other bandwidth selection techniques in the kernel smoothing literature, such as various \textit{plug-in} and \emph{biased} cross-validation methods (e.g. \cite{scoter:1987,shejon:1991,jonkap:1992}; see in addition the collection of references in Section 3.9 of Wand \& Jones \cite{wj:1995}); though many are not readily applicable to global bandwidth selection in adaptive bandwidth estimation, or indeed even for fixed bandwidth estimators when the data are spatially (and possibly temporally) bounded. Bayesian approaches to bandwidth selection for multivariate kernel density estimation have also been discussed \cite{zhaetal:2006,huetal:2012}, though we have had difficulty in replicating the reported success of such selection methods for highly heterogeneous spatial and spatiotemporal data, and further research is warranted.

Out of the methods we have discussed, we have found the bootstrap approach to perform well overall when using the oversmoothing bandwidth for construction of a reference density. Our concurrent research efforts are focussed in part on this type of smoothed cross-validation in order to better understand its performance and guide appropriate reference density estimation for the kinds of applications of interest.

\section{Spatial Relative Risk}\label{sec:srr}

The idea of the \emph{spatial relative risk function} was first proposed by Bithell \cite{bit:1990,bit:1991} as a way to investigate the relative abundance of disease cases with respect to the at-risk population dispersion over a well-defined geographical region. This was developed further in an epidemiological setting by Kelsall \& Diggle \cite{keldig:1995a,keldig:1995b}, who looked specifically at edge-corrected fixed bandwidth kernel estimators of the form of (\ref{eq:kdefix}) for the requisite densities. Adaptive kernel estimation of spatial relative risk was subsequently developed in Davies \& Hazelton \cite{davhaz:2010} and Davies et al. \cite{davetal:2016}. In this section we present an overview of the various estimators and methods of inference, focussing on practical aspects of their implementation.

\subsection{Ratio}\label{sec:ssrrat}
The estimator of spatial relative risk is a straightforward ratio of two kernel-estimated density functions defined on a common study window $W\subset\RR$. Suppose $\bX=\{\bx_1,\ldots,\bx_{n_1}\}$ and $\bY=\{\by_1,\ldots,\by_{n_2}\}$, the \emph{case} and \emph{control} data respectively, are two distinct samples of planar points assumed to originate from (unknown, possibly equivalent) density functions $f$ and $g$ respectively. Furthermore, let $\tilde{f}_{h_1}$ and $\tilde{g}_{h_2}$ be fixed bandwidth kernel estimates of the case and control densities $f$ and $g$. Their ratio $\tilde{r}=\tilde{f}_{h_1}/\tilde{g}_{h_2}$ is an estimate of the relative risk function $r=f/g$; more commonly expressed on the (natural) log scale as $\rho=\log f - \log g$ and estimated as
\begin{equation}\label{eq:rhofix}
\tilde{\rho}_{h_1,h_2}(\bx|\bX,\bY)=\log\big[\tilde{r}(\bx|\bX,\bY)\big]=\log\big[\tilde{f}_{h_1}(\bx|\bX)\big]-\log\big[\tilde{g}_{h_2}(\bx|\bY)\big];\qquad\bx\in W.
\end{equation}
A flat surface at $\tilde{\rho}\approxeq 0$ suggests $f\approxeq g$. Peaks in the surface $\tilde{\rho}>0$ suggest a higher localised concentration of cases relative to the controls in the affected spatial areas, where troughs $\tilde{\rho}<0$ indicate a relative lack of cases. More generally, for applications outside the epidemiological discipline, we may simply refer to $f$ and $g$ as the `numerator' and `denominator' densities respectively; the interpretation of their ratio being the same in terms of the samples the two densities represent.

In practice, it has been shown to be desirable to employ a common bandwidth for estimation of the case and control densities of (\ref{eq:rhofix}) i.e.\ use $h=h_1=h_2$. This is due to a cancellation of bias terms in the asymptotics of $\tilde{\rho}$ when $f\approxeq g$ (see \cite{keldig:1995a}), leading to approximate unbiasedness of the log-risk estimator in the affected areas of $W$. The issue of finding some bandwidth $h$ that is \emph{jointly} optimal given the samples of $\bX$ and $\bY$ is thus brought to bear (see the upcoming discussion in Section \ref{sec:jointh}). In this situation, (\ref{eq:rhofix}) is rewritten as $\tilde{\rho}_h(\bx|\bX,\bY)=\log\big[\tilde{f}_{h}(\bx|\bX)\big]-\log\big[\tilde{g}_{h}(\bx|\bY)\big]$.

When estimated using spatially adaptive kernel smoothing (\ref{eq:kdeada}), we obtain the following relative risk surface assuming a global bandwidth $h_0$ common to both case and control densities:
\begin{equation}\label{eq:rhoada}
\hat{\rho}_{h_0}(\bx|\bX,\bY)=\log\big[\hat{f}_{h_0}(\bx|\bX)\big]-\log\big[\hat{g}_{h_0}(\bx|\bY)\big];\qquad\bx\in W.
\end{equation}
The work in \cite{davhaz:2010} revealed there to be both theoretical and practical benefits to the adaptive relative risk surface $\hat{\rho}$ over the fixed bandwidth version of $\tilde{\rho}$, particularly so when the observed patterns in $\bX$ and $\bY$ depart strongly from uniformity.

Use of the adaptive density estimator presents additional practical considerations that are not immediately obvious from (\ref{eq:rhoada}). The main issue here involves specification of the pilot densities used in calculation of Abramson's \cite{abram:1982} variable bandwidths as per (\ref{eq:hu}). The default implementation is to base each of the two pilots on their respective data sets, implying separate fixed bandwidth pilot density estimates (possibly with different pilot bandwidths $\tilde{h}_1$ and $\tilde{h}_2$) of
\begin{equation}\label{eq:asypilot}
\tilde{f}_{\tilde{h}_1}(\bx|\bX)\qquad\textrm{and}\qquad\tilde{g}_{\tilde{h}_2}(\bx|\bY)
\end{equation}
in (\ref{eq:hu}) for subsequent calculation of $\hat{f}_{h_0}$ and $\hat{g}_{h_0}$ respectively. This is results in different variable bandwidth factors for the numerator and denominator adaptive density estimates, and the corresponding ratio is referred to as the \emph{asymmetric} adaptive relative risk function. An alternative is to force equivalency of these bandwidth factors, achieved by using the same pilot density in both $\hat{f}_{h_0}$ and $\hat{g}_{h_0}$ in implementation of (\ref{eq:rhoada}), thereby providing the \emph{symmetric} adaptive relative risk function \cite{davetal:2016}. The common pilot density can be reasonably based on the pooled data $\bX\cup\bY$, leading to the estimate $\tilde{c}_{\tilde{h}}$:
\begin{equation}\label{eq:sympilot}
\tilde{f}_{\tilde{h}}(\bx|\cdot)=\tilde{g}_{\tilde{h}}(\bx|\cdot)=\tilde{c}_{\tilde{h}}(\bx|\bX\cup\bY)\quad\forall\quad\bx\in W.
\end{equation}
In turn, this means the bandwidth functions used for each adaptive density estimate would both be written as become $h(\cdot|c)$, as opposed to separate quantities $h(\cdot|f)$ and $h(\cdot|g)$ as implied by (\ref{eq:kdeada}) and (\ref{eq:hu}). Davies et al. \cite{davetal:2016} demonstrate theoretically and empirically that the symmetric adaptive estimator can be preferable to the asymmetric version when we assume the true numerator density $f$ to be similar to the denominator density $g$ over much of $W$; a situation that occurs frequently in many applications.

\subsection{Jointly Optimal Bandwidths}\label{sec:jointh}
The presence of two samples, $\bX$ and $\bY$, coupled with the aforementioned appeal of using the same bandwidth for the corresponding density estimates, introduces the difficult problem of somehow using the data to inform choice of a jointly optimal bandwidth for both estimates. To date, attention has been focussed on the fixed bandwidth spatial log relative risk estimator as per (\ref{eq:rhofix}), for which three related approaches have been proposed. We label these JOI$_a(h|\bX,\bY)$, and the goal in each is to find
\[
h_{\textrm{JOI}_a}=\textrm{argmin}_h\left[\textrm{JOI}_a(h|\bX,\bY)\right];\qquad a=1,2,3.
\]

The first is due to Kelsall \& Diggle \cite{keldig:1995a}, which is based on approximate minimisation of the MISE of $\tilde{\rho}_{\tilde{h}}$ via leave-one-out LSCV, not unlike the standalone-density LSCV selector in Section \ref{sec:lscv}. The criterion is
\begin{equation}\label{eq:joi1}
\textrm{JOI}_1(h|\bX,\bY)=2n_2^{-1}\sum_{j=1}^{n_2}\frac{\tilde{\rho}_h(\by_j|\bX,\bY_{[-j]})}{\tilde{g}_h(\by_j|\bY_{[-j]})}-2n_1^{-1}\sum_{i=1}^{n_1}\frac{\tilde{\rho}_h(\bx_i|\bX_{[-i]},\bY)}{\tilde{f}_h(\bx_i|\bX_{[-i]})}-\int_W\tilde{\rho}_h(\bx|\bX,\bY)^2\dint\bx.
\end{equation}
Note the ratios of leave-one-out estimators appearing in the criterion above, with the log relative risk and separate density estimates evaluated in turn at each omitted case and control coordinate.

The second joint selector, also based on leave-one-out operations, was proposed by Hazelton \cite{haz:2008}. We minimise
\begin{equation}\label{eq:joi2}
\textrm{JOI}_2(h|\bX,\bY)=n_2^{-1}\sum_{j=1}^{n_2}\tilde{\rho}_h(\by_j|\bX,\bY_{[-j]})^2-2n_1^{-1}\sum_{i=1}^{n_1}\tilde{\rho}_h(\bx_i|\bX_{[-i]},\bY),
\end{equation}
which can be shown to minimise an approximation to a weighted MISE with respect to the control density; namely the quantity $\int_W \left\{\tilde{\rho}_h(\bx|\bX,\bY)-\rho(\bx)\right\}^2g(\bx)\dint\bx$. The rationale is to lend more weight to areas where we have more data when it comes to finding an optimal bandwidth; as such, this selector is in theory better suited to those problems where the numerator and denominator data sets appear to share a similar appearance over much of $W$.

A third suggestion was explored in Davies \cite{dav:2013} and is represented by
\begin{equation}\label{eq:joi3}
\textrm{JOI}_3(h|\bX,\bY,\psi)=h^{-2}A_1(h|\bX,\bY,\psi)+0.5h^4A_2(h|\bX,\bY,\psi).
\end{equation}
The quantities $A_1$ and $A_2$ are given by
\[
A_1(h|\bX,\bY,\psi)=n_1^{-1}\int_{\mB}\frac{\mR_h(\bx;K)}{\tilde{f}_h(\bx|\bX)}\dint\bx + n_2^{-1}\int_{\mB}\frac{\mR_h(\bx;K)}{\tilde{g}_h(\bx|\bY)}\dint\bx
\]
and
\[
A_2(h|\bX,\bY,\psi)=0.5\int_{\mB}\frac{\tilde{f}''_{\psi}(\bx|\bX)^2}{\tilde{f}_h(\bx|\bX)^{2}}\dint\bx-\int_{\mB}\frac{\tilde{f}''_\psi(\bx|\bX)\tilde{g}''_\psi(\bx|\bY)}{\tilde{f}_h(\bx|\bX)\tilde{g}_h(\bx|\bY)}\dint\bx+0.5\int_{\mB}\frac{\tilde{g}''_{\psi}(\bx|\bY)^2}{\tilde{g}_h(\bx|\bY)^{2}}\dint\bx,
\]
where $\psi>0$ is some constant bandwidth used for the estimates $\tilde{f}''_\psi$ and $\tilde{g}''_\psi$ (clarified momentarily), $\mB\subseteq W$ is a possibly reduced spatial window (computed with respect to $\psi$; for example, as a morphological erosion of $W$ by some suitable distance with respect to the coverage of the kernel $K$), and
\begin{equation}\label{eq:R}
\mR_h(\bx;K)=\left\{q_h(\bx|W)h\right\}^{-2}\int_W K\left(\frac{\bu-\bx}{h}\right)^2\dint\bu.
\end{equation}
This selector is based on a direct plug-in approach to estimating the asymptotic MISE of $\tilde{\rho}_h$. The expression contains particular kernel estimates $\tilde{f}''_\psi(\bx|\bX)=\nabla^2\tilde{f}_\psi(\bx|\bX)$, defined as the sum of the unmixed second partial derivatives of the kernel density with respect to coordinate (defined in the same way for $\tilde{g}''_\psi$). Evaluation over a reduced spatial window $\mB$ is owed to the inability to easily edge-correct the plug-in kernel density derivative estimates; for a Gaussian $K$, eroding $W$ by e.g.\ $3\psi$ in order to obtain $\mB$ could be considered reasonable. Though not dependent on leave-one-out operations, the above is a rather crude approach that requires largely subjective choices for key elements in implementation, such as in setting $\psi$.

The cross-validation selectors of Kelsall \& Diggle \cite{keldig:1995a} and Hazelton \cite{haz:2008} suffer from the same practical problems as the standalone-density LSCV and LIK selectors discussed in Section \ref{sec:banddens}, and thus can have a tendency to produce bandwidths that undersmooth to a certain extent. Numerical results in \cite{dav:2013} showed (\ref{eq:joi3}) to avoid undersmoothing and thus be capable of outperforming (\ref{eq:joi1}) and (\ref{eq:joi2}), though it is important to be wary of the aforementioned ad hoc nature of the plug-in selector.

In principle, the approximations to the MISE and weighted MISE that lead to (\ref{eq:joi1}) and (\ref{eq:joi2}) can be applied to select a jointly optimal $h_0$ for the spatially adaptive relative risk estimator. This is achieved by replacing instances of $\tilde{\rho}_h$, $\tilde{f}_h$ and $\tilde{g}_h$ in these two equations by $\hat{\rho}_{h_0}$, $\hat{f}_{h_0}$ and $\hat{g}_{h_0}$. However, one complication is the consideration of the pilot bandwidth(s) used in the numerator and denominator densities, which are set \emph{a priori}, and the effects of this specification on subsequent optimisation of the common global bandwidth $h_0$ are difficult to predict.

\subsection{Monte-Carlo Tolerance Contours}\label{sec:tolmc}
In making inference, a natural question to ask with respect to an estimated spatial relative risk function is whether departures from the ``null'' level of (log) risk at zero are severe enough to constitute statistical evidence against equality of the numerator and denominator densities. This question is posed at the coordinate level, where the direction of the departure to be tested is driven by the application of interest. For example, in geographical epidemiology it is often of interest to identify areas of $W$ associated with an elevated risk of infection; in which case for a given $\bx\in W$ we would formulate the hypotheses
\begin{align}\label{eq:hyp}
\textrm{H}_0: & \quad\rho(\bx) = 0\nonumber\\
\textrm{H}_1: & \quad\rho(\bx) > 0.
\end{align}
Lower-tailed tests follow naturally by redefining $\textrm{H}_1$ as $\rho(\bx) < 0$. The result is a $p$-value surface, $P$, over $W$, that we may superimpose upon a plot of a corresponding estimate of $\rho$ at some elected significance level(s) (e.g.\ $\alpha=0.1,0.05,0.01$) as \emph{tolerance contours}.

A natural way to form these $p$-value surfaces is via Monte-Carlo (MC) simulation of kernel-estimated risk functions, as discussed in \cite{keldig:1995a}. Let $\tilde{\rho}_h(\bx|\bX,\bY)$ be a fixed bandwidth estimate of the log-relative risk function of interest. The algorithm is as follows:
\begin{enumerate}
\item Pool the case and control location data; $\bX\cup\bY$.
\item Randomly sample $n_1$ points from the pooled data without replacement to represent the simulated cases. Denote these $\bX^{(i)}$, for the $i$th iteration. The remaining $n_2$ points, $\bY^{(i)}$ are used as the simulated controls.
\item Find and store $\tilde{\rho}_h(\bx|\bX^{(i)},\bY^{(i)})$, $\forall$ $\bx\in W$.
\item Repeat 2 and 3 over $N$ iterations.
\end{enumerate}
Following the above, the $p$-value surface according to (\ref{eq:hyp}) is found as the proportion of simulated risk estimates that equal or exceed the estimated risk from the observed data at each evaluation coordinate $\bx$; namely
\begin{equation}\label{eq:P}
P(\bx)=(N+1)^{-1}\left\{1+\sum_{i=1}^{N}\mathbbm{1}\left[\tilde{\rho}_h(\bx|\bX,\bY)\leq \tilde{\rho}_h(\bx|\bX^{(i)},\bY^{(i)})\right]\right\};\qquad\bx\in W,
\end{equation}
where $\mathbbm{1}[\,\cdot\,]$ is the indicator function. From (\ref{eq:P}), $p$-values for the lower-tailed test can be computed simply as $1-P(\bx)$. 

The MC approach is easy to understand and implement, and is applied exactly as-is for the adaptive relative risk estimator (both asymmetric and symmetric versions) by replacing all instances of $\tilde{\rho}_h$ above by $\hat{\rho}_{h_0}$. However, there is an obvious computational cost that can be excessive, especially so when employing adaptive smoothing. A subtler issue is the fact that the MC algorithm conditions solely on the observed case and control locations, meaning the variation in the randomisation test at locations $\bx$ far from observed data is informed less by the process(es) at hand and more by arbitrary details of the particular implementation, such as the tail properties of the kernel function. In part to address these two issues, an alternative technique based on asymptotic properties of the kernel density estimator has been suggested, the practical details of which follow below.

\subsection{Asymptotic Tolerance Contours}\label{sec:tolasy}
The asymptotic (ASY) approach to calculating $p$-value surfaces for a spatial relative risk function based on the fixed bandwidth estimator $\tilde{\rho}_h$ was proposed in \cite{hazdav:2009}. Under natural assumptions regarding the choice of bandwidth $h$, standard theory for kernel estimation of probability density functions states that at a given coordinate $\bx\in W$, the estimates $\tilde{f}_h(\bx|\bX)$ and $\tilde{g}_h(\bx|\bY)$ is asymptotically normal \cite{par:1962}. Provided these two estimates are both bounded away from zero, this implies the limiting distribution of the corresponding log risk at $\bx$, $\tilde{\rho}_h(\bx|\bX,\bY)$, is also normal. 

Combining this idea with the cancellation of the leading bias term of $\tilde{\rho}_h$ thanks to use of a common numerator-denominator bandwidth $h$ (mentioned in Section \ref{sec:ssrrat}), under $\textrm{H}_0$ of (\ref{eq:hyp}) we have an approximately unbiased standard normal test statistic in
\begin{equation}\label{eq:zfix}
\tilde{Z}(\bx)=\frac{\tilde{\rho}_h(\bx|\bX,\bY)}{\sqrt{\V\left[\tilde{\rho}_h(\bx|\bX,\bY)\right]}};\qquad\bx\in W,
\end{equation}
and it remains to find an implementable quantity for $\V\left[\tilde{\rho}_h(\bx|\bX,\bY)\right]$. Taking the edge-correction factors $q_h(\bx|W)$ into account, Hazelton \& Davies \cite{hazdav:2009} suggest using a plug-in version of the asymptotic variance of the log risk estimator:
\begin{equation}\label{eq:varfix}
\V\left[\tilde{\rho}_h(\bx|\bX,\bY)\right]\approx\frac{\mR_h(\bx;K)}{\tilde{c}_h(\bx|\bX\cup\bY)h^2}\left(n_1^{-1}+n_2^{-1}\right).
\end{equation}
Here, $\mR_h(\bx;K)$ is given in (\ref{eq:R}), and $\tilde{c}_h(\bx|\bX\cup\bY)$ is a plug-in kernel estimate via (\ref{eq:kdefix}) of the pooled data set. We use the latter instead of separate case and control plug-in density estimates because we assume they are equal under the null hypothesis. The corresponding (upper-tailed) $p$-value surface directly results as
\[
P(\bx)=1-\Phi\left[\tilde{Z}(\bx)\right]
\]
for the standard normal cumulative distribution function $\Phi[\,\cdot\,]$. The comments made in Section \ref{sec:tolmc} regarding manipulation of $P(\bx)$ to obtain the lower-tailed version of the test also applies here.

While the ASY $p$-value surface for the fixed bandwidth risk function depends upon the plugged-in estimate  $\tilde{c}_h$, we have found the performance of (\ref{eq:zfix}) to be preferable to the MC $p$-value surface in terms of numerical stability. See \cite{hazdav:2009} for further details.

The same motivations lead to the ASY $p$-value surfaces for an adaptive relative risk estimate, which were introduced in Davies \& Hazelton \cite{davhaz:2010}. The test statistics are given by 
\begin{equation}\label{eq:zada}
\hat{Z}(\bx)=\frac{\hat{\rho}_{h_0}(\bx|\bX,\bY)}{\sqrt{\V\left[\hat{\rho}_{h_0}(\bx|\bX,\bY)\right]}};\qquad\bx\in W,
\end{equation}
and the variance approximations used in the above differ slightly for the asymmetric version of the estimator and the symmetric version. For an asymmetric estimate $\hat{\rho}_{h_0}$ it can be shown that
\begin{equation}\label{eq:varadaasy}
\V\left[\hat{\rho}_{h_0}(\bx|\bX,\bY)\right]\equiv\V_{\textrm{A}}\left[\hat{\rho}_{h_0}(\bx|\bX,\bY)\right]\approx h_0^{-2}\left(\frac{\mS_K(\bx;f)\gamma_f}{n_1}+\frac{\mS_K(\bx;g)\gamma_g}{n_2}\right),
\end{equation}
where
\[
\mS_K(\bx;f)=\left\{ q_{h(\bx;f)}(\bx|W) h_0\right\}^{-2}\left[2\int_W K\left(\frac{\bu-\bx}{h(\bx;f)}\right)^2\dint\bu+\frac{1}{4}\int_W M\left(\frac{\bu-\bx}{h(\bx;f)}\right)^2\dint\bu\right],
\]
with $M(\bu)=2K(\bu)+u_1\frac{\partial K}{\partial u_1}+u_2\frac{\partial K}{\partial u_2}$; $\bu\equiv[u_1,u_2]^\top$; defined similarly for $\mS_K(\bx;g)$. In practice it is advisable to find some appropriate value $\gamma_{fg}$ to replace both $\gamma_f$ and $\gamma_g$ in (\ref{eq:hu}) and hence (\ref{eq:varadaasy}), without which we miss out on the benefits afforded by use of a common global bandwidth $h_0$. We use $\gamma_{fg}=\sqrt{\gamma_f\gamma_g}$.

For a symmetric estimate based on a pooled-data pilot density $c$, (estimated as $\tilde{c}_{\tilde{h}}$), instead of using (\ref{eq:varadaasy}) in (\ref{eq:zada}), the appropriate formula is
\begin{equation}\label{eq:varadasym}
\V_{\textrm{S}}\left[\hat{\rho}_{h_0}(\bx|\bX,\bY)\right]\approx h_0^{-2}\mS_K(\bx;c)\gamma_c\left(\frac{1}{n_1}+\frac{1}{n_2}\right).
\end{equation}

The ASY $p$-value surfaces for the adaptive relative risk function offer substantial computational savings over the MC option. We have noted that the natural imbalance in the bandwidth factors associated with the asymmetric estimator yields contours more sensitive to local fluctuations in risk and hence the specific choice of common global bandwidth $h_0$, while contours based on the symmetric estimator provide a more conservative spatial test. For these comparisons and further technical details, see \cite{davhaz:2010} and \cite{davetal:2016}.

\section{Code Break: Spatial Relative Risk with \texttt{sparr}}\label{sec:srrcode}
The PBC data presented in Section \ref{sec:mot} are provided in \verb|sparr| as \verb|pbc|, which is an object of class \verb|ppp| (\emph{p}lanar \emph{p}oint \emph{p}attern), defined in the \verb|spatstat| package. The object contains the spatial coordinates and identification of the cases and controls as the \verb|marks|, as well as the study window stored as a polygonal \verb|owin|. The reader is directed to the relevant \verb|spatstat| help files \verb|help(ppp.object)| and \verb|help(owin.object)| to learn more about these classes, and the data themselves are described in \verb|sparr|'s documentation by typing \verb|help(pbc)|. For creation of the data plot in Figure \ref{fig:pbcdata}, see the supplementary material.

Our goal is to estimate the spatial relative risk of PBC with respect to the sampled controls. We will produce both fixed and adaptive bandwidth kernel estimates thereof. First, ensure \verb|sparr| is loaded with a call to \verb|library("sparr")|. The code

\small
\begin{verbatim}
R> data(pbc)
R> pbccas <- split(pbc)$case
R> pbccon <- split(pbc)$control
\end{verbatim}
\normalsize
then loads the dataset, and splits it into separate objects containing the 761 cases (\verb|pbccas|) and 3020 controls (\verb|pbccon|) for convenience. 

\subsection{Fixed Bandwidth Risk Surfaces}
We will begin with a simple fixed bandwidth estimate of relative risk, $\tilde{\rho}_h$, as per (\ref{eq:rhofix}). The quickest way to calculate and plot such an estimate is simply to supply the raw case and control data, as \verb|ppp| objects, to the \verb|f| and \verb|g| arguments of the function \verb|risk|:

\small
\begin{verbatim}
R> risk(f=pbccas,g=pbccon,doplot=TRUE)
Estimating case and control densities...Done.
\end{verbatim}
\normalsize

\begin{figure}[htbp]
\centering
\includegraphics[width=0.6\textwidth]{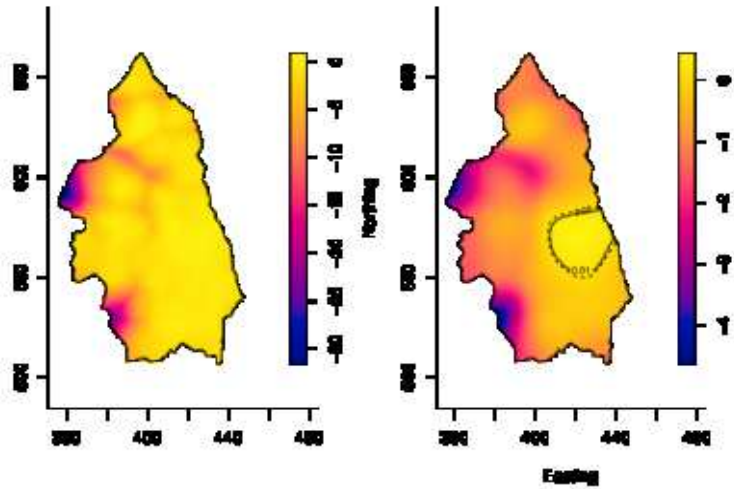}
\vspace{-4mm}
\caption{Fixed bandwidth spatial log-relative risk surfaces of the PBC data estimated using \texttt{sparr}. Left: A quick default estimate and plot obtained by supplying the raw case and control data directly to the \texttt{risk} function. Right: A recomputed estimate based on a jointly optimal, common case-control bandwidth found using \texttt{LSCV.risk}, as well as tolerance contours depicting significantly elevated risk of PBC calculated by MC simulations using \texttt{tolerance}.}\label{fig:pbcfix}
\end{figure}

If no bandwidth is supplied, the function internally computes $h_{\textrm{OS}}$ via (\ref{eq:hos}) using the pooled dataset for a common value, setting $n$ to be the geometric mean of the two sample sizes (cf.\ \cite{davhaz:2010}). A default image plot is produced with the optional argument \verb|doplot=TRUE|, and this is given on the left of Figure \ref{fig:pbcfix}. There is some visual indication of simultaneous over- and undersmoothing (a typical characteristic of fixed bandwidth smoothing of highly heterogeneous spatial data), the former in areas of dense observations, and the latter in the sparsely populated regions. At the very least, this suggests we should consider a more sophisticated choice of common bandwidth.

To do this, we turn to the jointly optimal selectors discussed in Section \ref{sec:jointh}, which are accessible via the \verb|LSCV.risk| function:

\small
\begin{verbatim}
R> hfix <- LSCV.risk(f=pbccas,g=pbccon)
Searching for optimal Kelsall-Diggle h in [0.1,15.278]...Done.
R> hfix
[1] 8.051627
\end{verbatim}
\normalsize

Supplying the case and control data to the \verb|f| and \verb|g| arguments as earlier, the default use of \verb|LSCV.risk| is to execute the JOI$_1$ selector, (\ref{eq:joi1}), of Kelsall \& Diggle \cite{keldig:1995a}. The JOI$_2$ and JOI$_3$ selectors can be requested via the optional \verb|method| argument. In this instance however, we note a common fixed bandwidth of around $h=8.05$ selected for these data, which we store as the object \verb|hfix|.

We now recompute the fixed bandwidth estimate of PBC risk, and choose to do so by first individually computing the numerator and denominator densities. Standalone spatial density estimation is achieved with the function \verb|bivariate.density|,

\small
\begin{verbatim}
R> f.tilde <- bivariate.density(pbccas,h0=hfix)
R> g.tilde <- bivariate.density(pbccon,h0=hfix)
\end{verbatim}
\normalsize
which by default computes a fixed bandwidth kernel estimate via (\ref{eq:kdefix}) with uniform edge correction (\ref{eq:qfix}). The desired isotropic smoothing bandwidth is supplied to \verb|h0|, to which in both instances we have passed the jointly optimal \verb|hfix| value calculated above. Calculation of the log-relative risk surface follows as before; the \verb|risk| argument also accepts objects of class \verb|bivden| returned by a call to \verb|bivariate.density|.

\small
\begin{verbatim}
R> rho.tilde <- risk(f=f.tilde,g=g.tilde)
\end{verbatim}
\normalsize
Here we have chosen to store the output as the object \verb|rho.tilde| rather than plot it straightaway using the \verb|doplot| argument. This is because we wish to answer the aforementioned research question concerning potentially significant departures of the risk from uniformity over the study region, and we shall do so by additionally computing appropriate $p$-value surfaces to ultimately plot tolerance contours.

The \verb|tolerance| function is used to this end, and is capable of computing both MC and ASY $p$-value surfaces as detailed in Sections \ref{sec:tolmc} and \ref{sec:tolasy} respectively, when supplied an object returned from a call to \verb|risk|. Its default behaviour is to compute ASY $p$-values (via (\ref{eq:zfix}) and (\ref{eq:varfix}) for a fixed bandwidth \verb|risk| object) though we will elect to compute the MC version by simulation to avoid the need to obtain the pooled density estimate $\tilde{c}_h$ as required for the fixed-bandwidth asymptotic variance in (\ref{eq:varfix}). The call

\small
\begin{verbatim}
R> pval.tilde <- tolerance(rs=rho.tilde,method="MC",ITER=200)
|=============================================================| 100%
\end{verbatim}
\normalsize
does just this; we supply the pre-computed risk surface object to \verb|rs|, and the argument \verb|method| is set to \verb|"MC"| for a Monte-Carlo $p$-value surface. The \verb|ITER| argument stipulates the number of iterations to run---this is represented by $N$ in (\ref{eq:P}). A dynamic progress bar displays completion; on the lead author's desktop machine these simulations took around 35 seconds.

The resulting object \verb|pval.tilde| is a pixel image (\verb|spatstat| class \verb|im|) of $p$-values corresponding to an upper-tailed test of increased risk, and we may superimpose corresponding tolerance contours atop an existing plot of the log-relative risk surface of interest. The code

\small
\begin{verbatim}
R> plot(rho.tilde,xlab="Easting",ylab="Northing")
R> tol.contour(pim=pval.tilde,levels=c(0.01,0.05),lty=1:2,add=TRUE)
\end{verbatim}
\normalsize
uses \verb|sparr|'s \verb|tol.contour| function to do this, displaying dashed- and solid-line contours upon the risk surface plot at significance levels of 0.05 and 0.01 respectively. The optional argument \verb|test| can be used to change the contours to show a lower-tailed test if desired, with remaining aesthetics of the contours themselves controlled by arguments passed to the built-in \verb|R| function \verb|contour|. This final result is given on the right of Figure \ref{fig:pbcfix}. In terms of visual appearance of the risk surface itself, it seems improved somewhat over the initial simple estimate, and the added tolerance contours shows an overall heightened area of risk of PBC along the eastern border covering much of that densely populated sub-region. This mirrors the results of the original analysis in Prince et al. \cite{prinetal:2001}.

\subsection{Adaptive Bandwidth Risk Surfaces}\label{sec:adacode}
The fact that the fixed bandwidth tolerance contours in Figure \ref{fig:pbcfix} do little to highlight any kind of detail in the densely populated sub-region is a natural consequence of the generous amount of smoothing applied to the risk function estimate. Given the relatively large amount of data in that area, it would be beneficial to see if we can tease out more structure in the estimated risk, particularly with respect to significantly heightened risk. While we could try to simply reduce the fixed bandwidth, this would result in instability in the remainder of the region where data are sparse, heightening the chances of `false-positive' contours. As discussed in Section \ref{sec:kdeada}, adaptive smoothing of a density estimate is designed to provide us with the flexibility of reduced smoothing in densely occupied areas so that structural detail can be captured, without compromising the stability of the estimate elsewhere. It is therefore worth assessing an adaptive kernel estimate of the spatial relative risk for these data.

An asymmetric adaptive estimate is obtained by applying (\ref{eq:kdeada}) with (\ref{eq:hu}) separately to the case and control samples as noted in (\ref{eq:asypilot}). First, we need to choose bandwidths for the pilot estimation stage, $\tilde{h}_1$ and $\tilde{h}_2$ in (\ref{eq:asypilot}), which we will do with the bootstrap selector defined by (\ref{eq:boot}). This is implemented as the \verb|BOOT.density| function; by executing

\small
\begin{verbatim}
R> hpilot.f <- BOOT.density(pbccas)
Initialising...Done.
Searching for optimal h in [0.0999999999999091, 15.2778333333333]...Done.
R> hpilot.f
[1] 2.499082
R> hpilot.g <- BOOT.density(pbccon)
Initialising...Done.
Searching for optimal h in [0.0999999999999091, 15.2778333333333]...Done.
R> hpilot.g
[1] 1.984562
\end{verbatim}
\normalsize
we obtain bandwidths $h_\textrm{BOOT}$ for the case (\verb|hpilot.f|) and control (\verb|hpilot.g|) pilot densities. By default, \verb|BOOT.density| uses the oversmoothing bandwidth $h_\textrm{OS}$ of (\ref{eq:hos}) for each reference bandwidth $\eta$.

Next, we will simply use the oversmoothing bandwidth calculated on the pooled dataset for a common numerator/denominator global bandwidth. The line

\small
\begin{verbatim}
R> h0 <- OS(pbc,nstar="geometric")
R> h0
[1] 3.498445
\end{verbatim}
\normalsize
does this via the \verb|OS| command; replacing $n=n_1+n_2$ in (\ref{eq:hos}) by the geometric mean of the case and control samples sizes $n=\sqrt{n_1n_2}$ (to gain some sensitivity for the density estimate that corresponds to the smaller sample size).

The asymmetric adaptive kernel log-relative risk function estimate can then be obtained with a direct call to \verb|risk| as follows.

\begin{figure}[hbpt]
\centering
\includegraphics[width=0.7\textwidth]{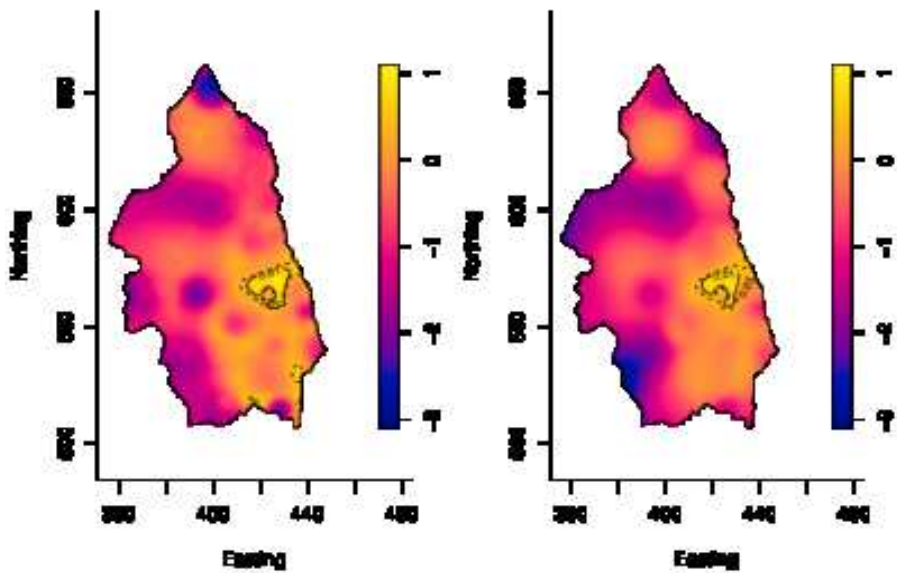}
\caption{Adaptive bandwidth spatial log-relative risk surfaces of the PBC data estimated using \texttt{sparr}, with ASY tolerance contours displayed on both. Left: An asymmetric estimate based on bootstrap-estimated pilot bandwidths and an oversmoothing global bandwidth calculated using the pooled data. Right: A symmetric estimate based on a pooled-data pilot density, using the same bandwidths.}\label{fig:pbcada}
\end{figure}

\small
\begin{verbatim}
R> rho.hat1 <- risk(f=pbccas,g=pbccon,h0=h0,adapt=TRUE,hp=c(hpilot.f,hpilot.g),tolerate=TRUE)
Estimating pilot(s)...Done.
Estimating case density...Done.
Estimating control density...Done.
Calculating tolerance contours...Done.
\end{verbatim}
\normalsize
Here, we have again supplied the raw data to \verb|risk|, but we have set the optional argument \verb|adapt=TRUE| to instruct the two density estimates to be computed as adaptive smooths via (\ref{eq:kdeada}) with (\ref{eq:hu}) and uniform edge correction (\ref{eq:qada}) (automatically using, as noted in Section \ref{sec:tolasy}, a common geometric mean scaler $\gamma_{fg}$ for equality of the overall global smoothing). The common global bandwidth itself is passed to \verb|h0|, and the two pilot bandwidths are provided in the order of (\textit{case}, \textit{control}) as a vector of length two to \verb|hp|. Finally, note the optional argument \verb|tolerate| has been set to \verb|TRUE|, which means \verb|risk| will also internally compute and return ASY $p$-value surfaces (via (\ref{eq:zada}) and (\ref{eq:varadaasy}) for the asymmetric adaptive version) for subsequent tolerance contour plotting. A plot, given on the left of Figure \ref{fig:pbcada}, is produced by executing

\small
\begin{verbatim}
R> plot(rho.hat1,zlim=c(-3.1,1.1),tol.args=list(levels=c(0.01,0.05),lty=1:2),
        xlab="Easting",ylab="Northing")
\end{verbatim}
\normalsize
In the call to \verb|plot|, we control the plotted limits of the surface through \verb|zlim|, and we use the optional \verb|tol.args| argument to instruct the function which levels of the pre-computed $p$-value surface in \verb|rho.hat1| to display as tolerance contours. The image shows a similar localisation of significantly increased risk of PBC as in the fixed bandwidth estimate, albeit with greater detail as we would expect given the reduced smoothing in that area. There is some indication of minor pockets of increased risk in the south at the 5\% level, though we are inclined to interpret these with caution, given both the small size of the encapsulated areas and their proximity to the boundary. 

We contrast the asymmetric adaptive log-risk surface with a symmetric version as detailed in Section \ref{sec:ssrrat}. We will do so using a pilot density estimate based on the pooled data, $\tilde{c}_{\tilde{h}}(\cdot|\bX\cup\bY)$ as noted in (\ref{eq:sympilot}), to calculate the bandwidth factors of (\ref{eq:hu}) for both case and control adaptive density estimates. Consider the following:

\small
\begin{verbatim}
R> hpilot.pool <- BOOT.density(pbc)
Initialising...Done.
Searching for optimal h in [0.0999999999999091, 15.2778333333333]...Done.
R> hpilot.pool
[1] 1.777469
R> f.hat <- bivariate.density(pbccas,h0=h0,hp=hpilot.pool,adapt=TRUE,pilot.density=pbc)
===================================================================
R> g.hat <- bivariate.density(pbccon,h0=h0,hp=hpilot.pool,adapt=TRUE,pilot.density=pbc)
===================================================================
\end{verbatim}
\normalsize
The first call to \verb|BOOT.density| finds a bandwidth for the pilot density estimate using the pooled, original data object \verb|pbc|. Then, we create case and control density estimates using \verb|bivariate.density|, setting \verb|adapt=TRUE|, storing the results as \verb|f.hat| and \verb|g.hat| respectively (note a progress bar is displayed by default for adaptive density estimates). In both estimates, we use the optional \verb|pilot.density| argument, and pass it the pooled dataset. This instructs \verb|bivariate.density| to base both sets of variable bandwidths on a pilot density (using the pre-computed pilot bandwidth \verb|hp=hpilot.pool|) constructed from the pooled data. Subsequently, the line

\small
\begin{verbatim}
R> rho.hat2 <- risk(f.hat,g.hat,tolerate=TRUE)
Calculating tolerance contours...Done.
\end{verbatim}
\normalsize
takes these two adaptive densities and returns a risk surface object with a corresponding $p$-value surface (this time, the ASY values are computed using (\ref{eq:zada}) with (\ref{eq:varadasym})). Then, the same call to \verb|plot| as above, replacing \verb|rho.hat1| with \verb|rho.hat2|, produces the image on the right of Figure \ref{fig:pbcada}.

In comparison to the asymmetric estimate, we observe a little more smoothness and stability, particularly in the less-populated areas; a noted characteristic of the symmetric adaptive estimator \cite{davetal:2016}, which is also less sensitive overall to minor changes to the global and pilot bandwidths. Importantly, the main message regarding the area of significantly elevated risk is much the same for both adaptive estimates. A particularly encouraging feature of the symmetric estimate for the PBC example is that it retains said smoothness in areas of low population much like the generously smoothed fixed bandwidth surface on the right of Figure \ref{fig:pbcfix}, but does not obliterate the detail in the densely populated area.

The functionality of \verb|sparr| for 2D density estimation and spatial relative risk extends beyond the brief snippets and function usages we have demonstrated here (refer to Appendix \ref{app:sparr} for key features), and there are often different ways to go about the same task. For example, the \verb|rho.hat2| object could also have been created without the need to use \verb|bivariate.density| as

\small
\begin{verbatim}
R> rho.hat2 <- risk(pbccas,pbccon,h0=h0,hp=hpilot.pool,adapt=TRUE,tolerate=TRUE,pilot.symmetry="pooled")
\end{verbatim}
\normalsize
The reader is encouraged to study the comprehensive help files accompany the functions of \verb|sparr| to learn more.

\section{Spatiotemporal Relative Risk}\label{sec:strr}

The ideas in Section \ref{sec:srr} extend naturally to the spatiotemporal domain. Definition and implementation of the spatiotemporal relative risk function includes unique practical considerations, including the concept of spatial relative risk conditional on a specific time and the combination of a purely spatial density with one defined in space-time to assess problems where one distribution is assumed static in time. These ideas were discussed in detail by Fernando \& Hazelton \cite{ferhaz:2014}, and we provide an overview of the relevant methodology in this section.

\subsection{Time-Varying Denominator}\label{sec:tvden}
Like in the purely spatial domain, the idea of relative risk in space-time is based on estimation of the ratio of two density function defined on appropriately bounded intervals. Using notation from Section \ref{sec:st}, we assume the presence of two data samples, each containing the spatial (in 2D) and temporal (in 1D) event locations. Denote these $\bmX=\{(\bx_1,t_1),\ldots,(\bx_n,t_{n_1})\}$ for $n_1$ `numerator' or `case' observations, and $\bmY=\{(\by_1,u_1),\ldots,(\by_n,u_{n_2})\}$ for $n_2$ `denominator' or `control' observations, with spatial and temporal bounding regions $W\subset\RR$ and $T\subset\R$ such that $\bmX,\bmY\in W\times T\subset\RRR$. The goal is estimation of the spatiotemporal densities $f$ and $g$ assumed to have generated $\bmX$ and $\bmY$ respectively, in order to obtain (as an estimate of the true log-risk function $\rho=f/g$) the corresponding \emph{joint} log-risk surface
\begin{equation}\label{eq:rrst}
\breve{\rho}_{h,\lambda}(\bz,s|\bmX,\bmY)=\log\left[\breve{f}_{h,\lambda}(\bz,s|\bmX)\right]-\log\Big[\breve{g}_{h,\lambda}(\bz,s|\bmY)\Big];\qquad\bz\in W;\quad s\in T,
\end{equation}
where $\breve{f}_{h,\lambda}$ and $\breve{g}_{h,\lambda}$ are spatiotemporal estimates obtained via (\ref{eq:kdest}). Note the familiar use of a common set of smoothing bandwidths for both densities, $h$ in the spatial margin and $\lambda$ in the temporal margin; it is straightforward to show that the aforementioned asymptotic bias reduction in areas of equivalent numerator and denominator density also applies in this setting.

The idea of conditioning on a specific time point $t$ is expressed via the \emph{conditional} spatiotemporal relative risk surface,
\begin{equation}\label{eq:rrstcond}
\breve{\rho}_{h,\lambda}(\bz|s=t,\bmX,\bmY)=\log\left[\breve{f}_{h,\lambda}(\bz|s=t,\bmX)\right]-\log\Big[\breve{g}_{h,\lambda}(\bz|s=t,\bmY)\Big];\qquad\bz\in W;\quad t\in T,
\end{equation}
where $\breve{f}_{h,\lambda}(\bx|s=t,\bmX)$ is an estimate of spatial numerator density at time $t$, given by (\ref{eq:kdestcond}); defined similarly for the denominator density. As noted in Section \ref{sec:st}, the difference between the joint and conditional versions of the estimator is simply down to normalisation, with the latter estimator offering a 2D, spatial-only snapshot of relative risk, and the former, if plotted at time $t$, merely represents a unnormalised `slice' of the 3D function. The relationship expressed by (\ref{eq:kdestcond}) means (\ref{eq:rrstcond}) can be obtained directly from (\ref{eq:rrst}) as
\[
\breve{\rho}_{h,\lambda}(\bz|s=t,\bmX,\bmY)=\breve{\rho}_{h,\lambda}(\bz,t|\bmX,\bmY)+\log\Big[\bar{g}_\lambda(t|\bu)\Big]-\log\Big[\bar{f}_\lambda(t|\bt)\Big],
\]
where $\bar{f}_\lambda$ and $\bar{g}_\lambda$ are univariate density estimates of the temporal margins of the two samples as per (\ref{eq:ft}); here we have $\bt=\{t_1,\ldots,t_{n_1}\}$ and $\bu=\{u_1,\ldots,u_{n_2}\}$.

\subsection{Time-Constant Denominator}\label{sec:tcden}
The estimator of spatiotemporal relative risk changes if we consider the situation where one set of observations is only observed in space; that is, when one of the densities on $W$ is static or constant over time. This is common in epidemiology, where the at-risk population distribution remains the same, but the distribution of infected individuals is prone to change during study period. As such, we consider explicitly the situation where the numerator density is estimated from a spatiotemporal data sample $\bmX=\{(\bx_1,t_1),\ldots,(\bx_n,t_{n_1})\}$, but the denominator density is based on a purely spatial dataset $\bY=\{\by_1,\ldots,\by_{n_2}\}$. The joint spatiotemporal log relative risk estimator of (\ref{eq:rrst}) thus becomes
\begin{equation}\label{eq:rrst2}
\breve{\rho}_{h,\lambda}(\bz,s|\bmX,\bY)=\log\left[\breve{f}_{h,\lambda}(\bz,s|\bmX)\right]+\log\Big[|T|\Big]-\log\Big[\tilde{g}_h(\bz|\bY)\Big];\qquad\bz\in W;\quad s\in T,
\end{equation}
where $\breve{f}_{h,\lambda}$ is obtained from (\ref{eq:kdest}) as before, $\tilde{g}_h$ is a purely spatial density estimate via (\ref{eq:kdefix}), and $|T|$ denotes the length of the temporal interval $T$. The corresponding conditional spatiotemporal log relative risk is found as 
\begin{equation}\label{eq:rrstcond2}
\breve{\rho}_{h,\lambda}(\bz|s=t,\bmX,\bY)=\log\left[\breve{f}_{h,\lambda}(\bz,t|\bmX)\right]-\log\Big[\bar{f}_\lambda(t|\bt)\Big]-\log\Big[\tilde{g}_h(\bz|\bY)\Big];\qquad\bz\in W;\quad t\in T.
\end{equation}

\subsection{Bandwidth Specification}
While bandwidth selection methods for standalone spatiotemporal density functions inclusive of edge correction factors can usually be extended in a straightforward fashion from the spatial-only setting (cf.\ Section \ref{sec:banddens}), the issue of purely data-driven, jointly optimal bandwidth selection for a spatiotemporal log relative risk estimate is an area that has not yet received any substantial attention in the literature. In the scenario where both numerator and denominator densities are time-varying i.e.\ when the estimator of interest is (\ref{eq:rrst}), Fernando \& Hazelton \cite{ferhaz:2014} suggest a natural extension of the LSCV selector in (\ref{eq:joi1}), where we minimise the criterion
\begin{equation}\label{eq:joi4}
\textrm{JOI}_4(h,\lambda|\bmX,\bmY)=2n_2^{-1}\sum_{j=1}^{n_2}\frac{\breve{\rho}_h(\by_j,u_j|\bmX,\bmY_{[-j]})}{\breve{g}_h(\by_j,u_j|\bmY_{[-j]})}-2n_1^{-1}\sum_{i=1}^{n_1}\frac{\breve{\rho}_h(\bx_i,t_i|\bmX_{[-i]},\bmY)}{\breve{f}_h(\bx_i,t_i|\bmX_{[-i]})}-\int_W\int_T\breve{\rho}_h(\bx,t|\bmX,\bmY)^2\dint t\dint \bx
\end{equation}
simultaneously with respect to $h$ and $\lambda$. However, the authors report poor performance of this selector in practice. This is somewhat unsurprising given the numerical instability of the 2D version of this approach as noted in Section \ref{sec:banddens}; the increase in dimension is likely to exacerbate the problems associated with taking ratios of leave-one-out estimates at isolated point locations in space-time.  

A further complication arises when considering applications with time-static control densities noted above. In this instance, the extension (\ref{eq:joi4}) is not applicable because only one set of data possess timestamps---evaluation of the spatiotemporal density $\breve{f}_{h,\lambda}$ at purely spatial control locations $\bY$ is not possible. This transdimensional problem would therefore also affect any related selectors, such as the extension of the weighted MISE selector given in (\ref{eq:joi2}).

These difficulties aside, successful exploratory analysis of spatiotemporal relative risk is readily carried out by even subjective choices of $h$ and $\lambda$ (see examples in both \cite{ferhaz:2014} and \cite{zhaetal:2011}). A researcher can also appeal to the standalone spatiotemporal density versions of LSCV, LIK, and BOOT selectors detailed in Section \ref{sec:banddens}, possibly pooling the data $\bmX$ and $\bmY$ (or the spatial margin of $\bmX$ with $\bY$ in the time-constant denominator scenario). For a quick initial estimate of $h$ and $\lambda$, we find that the 2D and 1D versions of the oversmoothing bandwidth as per (\ref{eq:hosd}) form an excellent starting point in practice.

\subsection{Tolerance Contours}\label{sec:sttol}
Just as in testing for anomalous sub-regions in spatial relative risk surfaces, the same type of questions may be posed in the context of space-time. We do this with respect to either the joint relative risk
\begin{align}\label{eq:hypstj}
\textrm{H}_0: & \quad\rho(\bz,s) = 0\nonumber\\
\textrm{H}_1: & \quad\rho(\bz,s) > 0;
\end{align}
or the conditional relative risk
\begin{align}\label{eq:hypstc}
\textrm{H}_0: & \quad\rho(\bz|s=t) = 0\nonumber\\
\textrm{H}_1: & \quad\rho(\bz|s=t) > 0.
\end{align}
While we pose the above hypotheses as an upper-tailed test for significantly elevated risk in space and time; as before, a lower-tailed test is also theoretically valid.

In order to test hypotheses such as (\ref{eq:hypstj}) and (\ref{eq:hypstc}), MC simulations based on random permutations of the case and control labels on the pooled dataset can be utilised exactly as outlined in Section \ref{sec:tolmc}. However, computational expense is even more noticeable in the spatiotemporal setting due to the increase in dimension, and the aforementioned drawbacks of conditioning solely on the observed coordinates remain. ASY $p$-value surfaces are therefore particularly convenient for this type of testing. The key premise is the same as earlier---we use a plug-in version of the appropriate asymptotic variance to compute an approximately normal test statistic at $(\bz,s)$.

For a spatiotemporal relative risk estimate based on two time-varying densities (Section \ref{sec:tvden}), the test statistics are found as either
\begin{equation}\label{eq:zstjc}
\breve{Z}(\bz,s)=\frac{\breve{\rho}_{h,\lambda}(\bz,s|\bmX,\bmY)}{\sqrt{\V\left[\breve{\rho}_{h,\lambda}(\bz,s|\bmX,\bmY)\right]}}\qquad\textrm{or}\qquad \breve{Z}(\bz|s=t)=\frac{\breve{\rho}_{h,\lambda}(\bz|t,\bmX,\bmY)}{\sqrt{\V\left[\breve{\rho}_{h,\lambda}(\bz|t,\bmX,\bmY)\right]}};\qquad\bz\in W;\quad s,t\in T,
\end{equation}
corresponding to the two sets of hypotheses in (\ref{eq:hypstj}) and (\ref{eq:hypstc}) respectively. The variances are approximated as
\begin{equation}
\V\left[\breve{\rho}_{h,\lambda}(\bz,s|\bmX,\bmY)\right]\approx \frac{\mR_h(\bz;K)\mR_\lambda(s;L)}{\breve{c}_{h,\lambda}(\bz,s|\bmX\cup\bmY)h^2\lambda}\left(n_1^{-1}+n_2^{-1}\right)
\end{equation}
for the joint and 
\begin{equation}
\V\left[\breve{\rho}_{h,\lambda}(\bz|t,\bmX,\bmY)\right]\approx \frac{\mR_h(\bz;K)\mR_\lambda(t;L)}{\breve{c}_{h,\lambda}(\bz|t,\bmX\cup\bmY)h^2\lambda}\left(\frac{1}{n_1\bar{f}_\lambda(t|\bt)}+\frac{1}{n_2\bar{g}_\lambda(t|\bu)}\right)
\end{equation}
for the conditional, where $\mR_h(\bz;K)$ is given in (\ref{eq:R}); $\mR_\lambda(t|L)$ is defined similarly as 
\[
\mR_\lambda(t;L)=w_\lambda(t|T)^{-2}\lambda^{-1}\int_T L\left(\frac{s-t}{\lambda}\right)^2\dint s;
\]
the $\bar{f}_\lambda$ and $\bar{g}_\lambda$ are the univariate temporal-margin densities obtained via (\ref{eq:ft}); and $\breve{c}_{h,\lambda}$ represents a spatiotemporal density estimate of the pooled case-control data.

The formulae change slightly if our relative risk surface has been calculated with one spatial density held constant over time. For a time-static control as defined in Section \ref{sec:tcden}, the joint and conditional test statistics are defined as
\begin{equation}\label{eq:zstjc2}
\breve{Z}(\bz,s)=\frac{\breve{\rho}_{h,\lambda}(\bz,s|\bmX,\bY)}{\sqrt{\V\left[\breve{\rho}_{h,\lambda}(\bz,s|\bmX,\bY)\right]}}\qquad\textrm{or}\qquad \breve{Z}(\bz|s=t)=\frac{\breve{\rho}_{h,\lambda}(\bz|t,\bmX,\bY)}{\sqrt{\V\left[\breve{\rho}_{h,\lambda}(\bz|t,\bmX,\bY)\right]}};\qquad\bz\in W;\quad s,t\in T.
\end{equation}
In this setting, it can be shown that the plug-in joint and conditional variances for the expressions in (\ref{eq:zstjc2}) are equivalent, and we use
\begin{equation}\label{eq:varasyst}
\V\left[\breve{\rho}_{h,\lambda}(\bz,s|\bmX,\bY)\right]=\V\left[\breve{\rho}_{h,\lambda}(\bz|s=t,\bmX,\bY)\right]\approx \frac{\mR_h(\bz;K)\mR_\lambda(s;L)}{\breve{f}_{h,\lambda}(\bz,s|\bmX)h^2n_1\lambda} + \frac{\mR_h(\bz;K)}{\tilde{g}_{h}(\bz|\bY)h^2n_2}.
\end{equation}
Full details of the above can be found in \cite{ferhaz:2014}.

\section{Code Break: Spatiotemporal Relative Risk with \texttt{sparr}}\label{sec:strrcode}
We turn our attention to the anonymised FMD data displayed in Figure \ref{fig:fmddata}, which has been made available in \verb|sparr| (courtesy of the Animal and Plant Health Agency (APHA), UK). For details, consult the documentation by executing \verb|help(fmd)|. Our goal is to use the methods detailed in Section \ref{sec:strr} to estimate the spatiotemporal relative risk of infection, and flag extreme localisations of spatial risk as they occur over the study period. As before, we shall load the dataset and for convenience extract the case and control data as separate objects:

\small
\begin{verbatim}
R> data(fmd)
R> fmdcas <- fmd$cases
R> fmdcon <- fmd$controls
\end{verbatim}
\normalsize
As the documentation states, these two objects are of \verb|spatstat| class \verb|ppp|, with the observation times of the infected cases stored as the \verb|marks| component of \verb|fmdcas|.

\subsection{Bandwidth Experimentation}
It is already clear from the brief description of the data in Section \ref{sec:mot} that the problem scenario is that of one described in Section \ref{sec:tcden}. That is, our numerator density is defined in space-time, but our denominator density only varies spatially. The issue of bandwidth selection is therefore somewhat complicated, with no directly available jointly optimal selectors spanning the two differently-dimensioned datasets. However, we can be guided by the results of appropriate standalone density selectors applied separately to the cases and controls, as well as the general preference to use a common spatial bandwidth $h$.

All bandwidth selectors noted in Section \ref{sec:banddens} have been implemented in \verb|sparr| for spatiotemporal data, with function names ending in ``\verb|spattemp|''. Examining the case data, we can find the 2D and 1D versions of the oversmoothing bandwidth (\ref{eq:hosd}) for a rough idea of appropriate spatial and temporal bandwidths $h$ (\verb|h|) and $\lambda$ (\verb|lambda|) with the call

\small
\begin{verbatim}
R> OS.spattemp(fmdcas)
      h   lambda 
5.86446 19.97246 
\end{verbatim}
\normalsize
Note that we do not explicitly need to supply the case observation times to this function, which will by default look to the \verb|marks| component of the object it is passed.

The corresponding spatial-only OS bandwidth for the controls is

\small
\begin{verbatim}
R> OS(fmdcon)
[1] 6.084745
\end{verbatim}
\normalsize
which is roughly the same as the $h$ for the cases.

Use, therefore, of a common $h$ of around $6$, and the above value of $\lambda$ for smoothing the temporal margin of the cases, would be a good starting point for estimation of the spatiotemporal relative risk of FMD. However, to avoid the adverse effects of potentially smoothing too excessively, it is also worth experimenting with more sophisticated selectors.

The likelihood CV selector described in Section \ref{sec:hlik}, for example, is implemented as the \verb|LIK.density| (spatial only) and \verb|LIK.spattemp| (spatiotemporal) functions. Executing the latter on the case dataset, we see the following:

\small
\begin{verbatim}
R> hlam <- LIK.spattemp(fmdcas)
h = 5.86446; lambda = 5.535485 
h = 6.450906; lambda = 5.535485 
h = 5.86446; lambda = 6.121931 
    << output suppressed >>
R> hlam
       h   lambda 
2.918775 8.828410 
\end{verbatim}
\normalsize
The function prints progress of the optimisation of the spatiotemporal version of (\ref{eq:lik}) with respect to $h$ and $\lambda$ as it runs (suppressed above solely for print). Edge-correction of both the spatial and temporal margins is included by default. The resulting bandwidths (stored as the object \verb|hlam|) are roughly half the size of those selected by the call to \verb|OS.spattemp|. Comparing the selected $h$ with that arrived at for the spatial-only controls,

\small
\begin{verbatim}
R> LIK.density(fmdcon)
Searching for optimal h in [0.0316163698317011, 13.0346107692943]...Done.
[1] 1.484514
\end{verbatim}
\normalsize
we see the latter is smaller still, which is in part expected due to the much larger control sample.

At this point, we make the decision to proceed using the case-selected spatial bandwidth $h=2.919$ for both case and control samples (alongside the value $\lambda=8.828$ for the temporal margin of the cases). While this may not be optimal for a control-only density, note that it is still considerably smaller than the control OS bandwidth indicated above. This, coupled with bias cancellations in areas where $f\approxeq g$ when a common $h$ is used, means concerns over smoothing too generously in the denominator density are mitigated to a certain extent.

\subsection{Density Estimation}
We may now proceed to density estimation as outlined in Section \ref{sec:tcden}. Estimation of a spatiotemporal density $\breve{f}_{h,\lambda}$ as per (\ref{eq:kdest}) is achieved with the \verb|spattemp.density| function. Consider the following:

\small
\begin{verbatim}
R> range(marks(fmdcas))
[1]  20 220
R> f.breve <- spattemp.density(fmdcas,h=hlam[1],lambda=hlam[2],tlim=c(10,230))
Calculating trivariate smooth...Done.
Edge-correcting...Done.
Conditioning on time...Done.
\end{verbatim}
\normalsize
By default, \verb|spattemp.density| will compute the density estimate by first setting the temporal bounds of $T$ to the range of the observed timestamps, and evaluate it at all integer times within these limits (for datasets that span a very wide $T$, it is prudent to instead set the temporal-axis resolution using \verb|tres|). In our case, note that we manually widen the time limits a little to $[10,230]$ (note from the first line that the data are observed on $[20,220]$) using the optional argument \verb|tlim|. As earlier, the function looks to the \verb|marks| component of the provided data object for the observation times themselves; the user can alternatively supply a numeric vector of equal length to the number of observations to the \verb|tt| argument. Edge-correction is automatically performed on both the spatial and temporal margins via (\ref{eq:qfix}) and (\ref{eq:wt}) respectively.

Objects returned from a call to \verb|spattemp.density| are of class ``\verb|stden|'', which, as well as the trivariate estimate itself, contains various components describing the smooth. We can immediately view kernel estimates of the spatial (ignoring time) and temporal (ignoring space) margins of the cases, stored as the pixel images \verb|spatial.z| and \verb|temporal.z|, with the lines 

\begin{figure}[hbtp]
\centering
\includegraphics[width=1\textwidth]{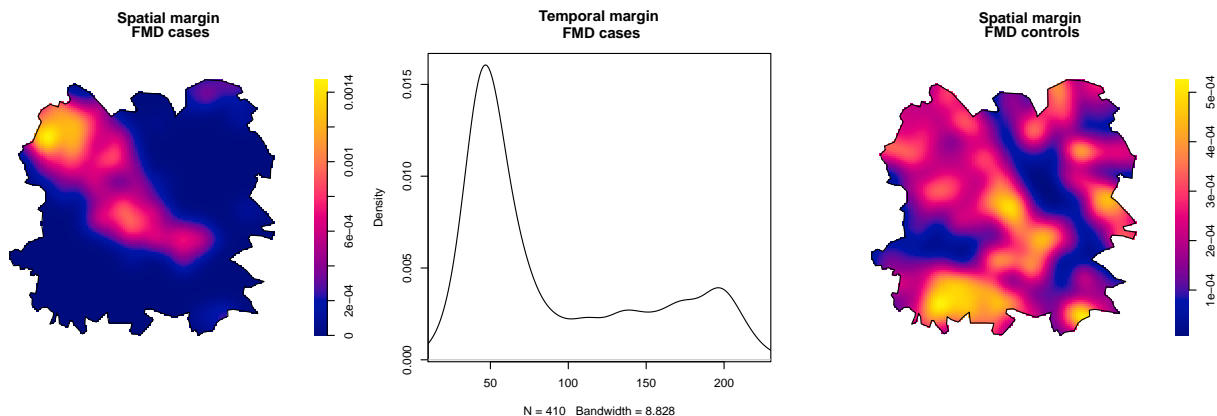}
\caption{Kernel estimates of the marginal spatial and temporal densities of the FMD cases (left and middle respectively), and the density estimate of the controls (right).}\label{fig:fmdden}
\end{figure}

\small
\begin{verbatim}
R> plot(f.breve$spatial.z,box=FALSE,main="Spatial margin\n FMD cases")
R> plot(Window(fmdcas),add=TRUE)
R> plot(f.breve$temporal.z,xlim=f.breve$tlim,xaxs="i",main="Temporal margin\n FMD cases")
\end{verbatim}
\normalsize
which produce the images on the left and middle of Figure \ref{fig:fmdden}. As we would expect, these marginal density estimates reflect the heterogeneity observed in the raw data on the left and middle of Figure \ref{fig:fmddata}.

The fixed bandwidth control density estimate, $\tilde{g}_h$, is provided by \verb|bivariate.density| (seen in Section \ref{sec:srrcode}) with the following simple line,

\small
\begin{verbatim}
R> g.tilde <- bivariate.density(fmdcon,h0=hlam[1])
\end{verbatim}
\normalsize
with a plot thereof created by 

\small
\begin{verbatim}
R> plot(g.tilde$z,box=FALSE,main="Spatial margin\n FMD controls")
R> plot(Window(fmdcon),add=TRUE)
\end{verbatim}
\normalsize
shown on the right of Figure \ref{fig:fmdden}.

\subsection{Relative Risk}\label{sec:strrsparr}
We are now in the position to create the spatiotemporal relative risk function estimate $\breve{\rho}_{h,\lambda}$ as per (\ref{eq:rrst2}), achieved with a call to \verb|spattemp.risk|:

\small
\begin{verbatim}
R> rho.breve <- spattemp.risk(f=f.breve,g=g.tilde,tolerate=TRUE)
Calculating ratio...Done.
Ensuring finiteness...
   --joint--
   --conditional--
Done.
Calculating tolerance contours...
   --convolution 1--
   --convolution 2--
Done.
\end{verbatim}
\normalsize
The function accepts precomputed objects of class \verb|stden| as the the case density to argument \verb|f|, and objects of either class \verb|stden| (for a time-varying denominator as per Section \ref{sec:tvden}) or \verb|bivden| (for a time-static denominator as per Section \ref{sec:tcden}) as the control density to argument \verb|g|; yielding the estimates defined by (\ref{eq:rrst}) and (\ref{eq:rrst2}) respectively. We additionally request calculation of ASY $p$-value surfaces defined, depending on scenario, by equations (\ref{eq:zstjc})-(\ref{eq:varasyst}), by stipulating \verb|tolerate=TRUE|.

At this point, the reader is encouraged to simply call

\small
\begin{verbatim}
R> plot(rho.breve)
\end{verbatim}
\normalsize
which plays the sequence of temporal slices of the joint relative risk estimate (\ref{eq:rrst2}) as an animation. By default, contours corresponding to an upper-tailed test at the 5\% level are superimposed. It is interesting to observe the movement of the epidemic over time with respect to the significantly elevated risk. Various options for tailoring these animations are available; see \verb|help(plot.rrst)|. 

The user can slice out the spatial margin at any desired selection of times. The \verb|spattemp.slice| returns both the corresponding spatial slices of the joint spatiotemporal risk surface (equations (\ref{eq:rrst}) and (\ref{eq:rrst2}) for a time-varying and time-constant denominator respectively); and the spatial relative risk surfaces conditional upon the specified times (equations (\ref{eq:rrstcond}) and (\ref{eq:rrstcond2}) for a time-varying and time-constant denominator respectively). Recall the difference between the joint and conditional versions of a spatiotemporal function in terms of normalisation as detailed in Sections \ref{sec:st} and \ref{sec:strr}; see also \cite{ferhaz:2014}. For example, the code

\small
\begin{verbatim}
R> mytimes <- c(20,40,60,90,150,200)
R> rho.breve.slices <- spattemp.slice(rho.breve,tt=mytimes)
R> names(rho.breve.slices)
[1] "rr"      "rr.cond" "P"       "P.cond" 
\end{verbatim}
\normalsize
sets up a vector of six desired times at which to return the spatial surfaces; these are supplied to \verb|tt| in the call to \verb|spattemp.slice|. The result is a named list of the selected joint and conditional spatial relative risk surfaces as the \verb|rr| and \verb|rr.cond| components respectively. Each of these members is itself a list of pixel images in the order corresponding to the times in \verb|tt|. Note in this instance we also have two additional members, \verb|P| and \verb|P.cond|, present because \verb|tolerate| was \verb|TRUE| in the original call to \verb|spattemp.risk|. These members provide the corresponding $p$-value surfaces to the pixel images in \verb|rr| and \verb|rr.cond|. 

\begin{figure}[hbpt]
\centering
\includegraphics[width=1\textwidth]{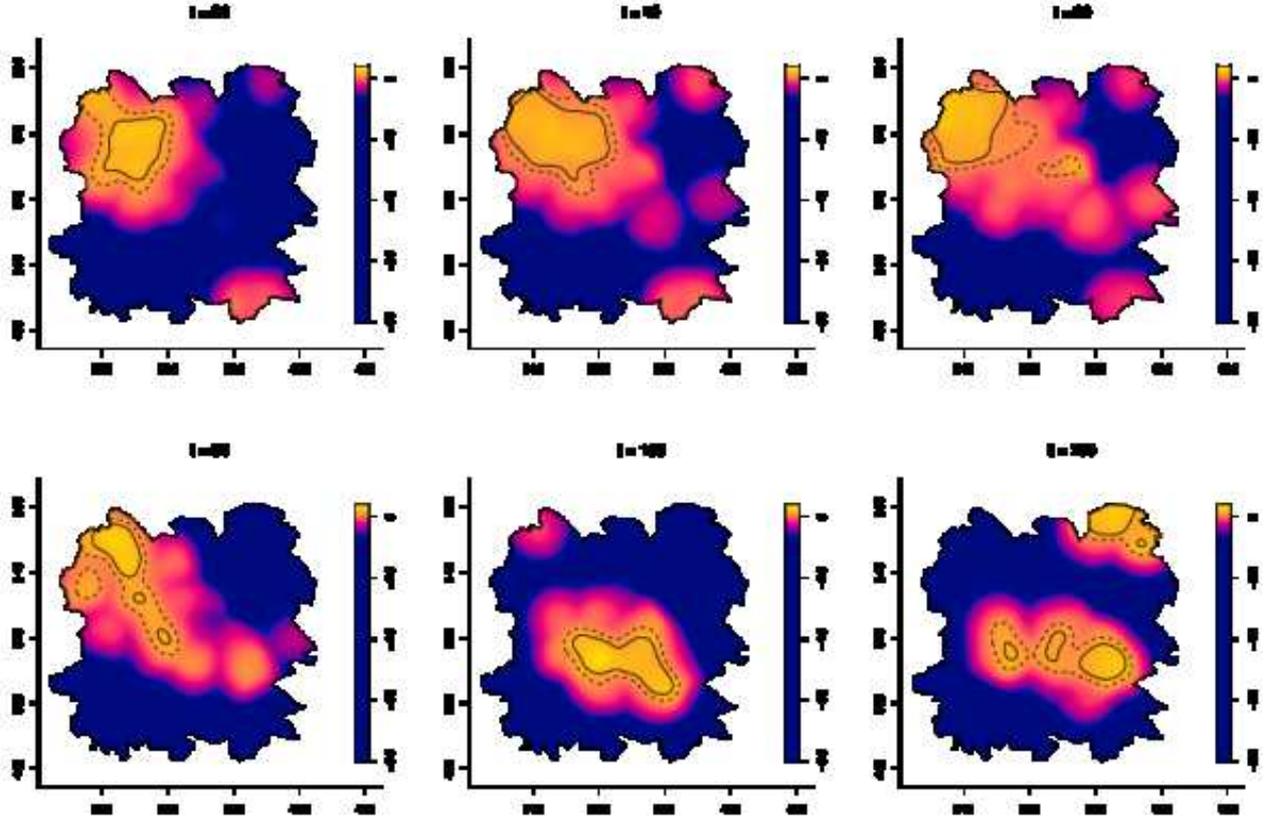}
\vspace{-10mm}
\caption{Relative risk of FMD farm infection in Cumbria conditional upon six specific times during the study period. Dashed contours flag statistical significance of elevated spatial risk at the 5\% level, and solid lines delinate a stringent 0.01\% level.}\label{fig:fmdcrr}
\end{figure}

The user can manually plot the images returned by \verb|spattemp.slice| for full control (see \verb|help(plot.im)| of the \verb|spatstat| package), or use the customisation available in \verb|sparr|'s \verb|plot.rrst| function shown above. A panelled series of plots of the conditional spatial relative risk functions, given the six times in \verb|mytimes|, appears in Figure \ref{fig:fmdcrr}; we defer the code that produces this to the supplementary material. From these plots we see the significantly heightened risk in the northwest in the early part of the study period moving south- south-east at later dates; trends visible in the earlier animation of the joint relative risk.

\section{Advanced Topics}\label{sec:adv}
To round off this tutorial, we will briefly discuss some additional topics that, while somewhat more technical, are directly relevant to the practical aspects of kernel estimation of spatial and spatiotemporal relative risk. These concern the computational complexities of spatially adaptive smoothing, and some advanced plotting techniques for multidimensional functions. Short demonstrations are provided in \verb|R| to motivate further experimentation by the reader.

\subsection{Partitioned and Multi-scale Adaptive Density/Intensity Estimation}\label{sec:multi}
By running the example code in Section \ref{sec:srrcode}, the reader will note the additional time it takes to compute the adaptive estimates over the fixed bandwidth estimates. The reason for this is due to the fact that the fixed bandwidth kernel density estimator is represented mathematically as a 2D \emph{convolution}, which allows for fast and efficient calculation thanks to a special numerical transformation referred to as the \emph{discrete Fourier transform} (DFT), thereby avoiding direct evaluation of (\ref{eq:kdefix}). Details of this computational short-cut, which represents the current standard for implementation of the traditional fixed bandwidth kernel estimator, can be found in Wand \cite{wand:1994}.

The same type of 2D convolution cannot be written down, however, if we allow the smoothing bandwidth to vary by data point; exactly what the adaptive estimator (\ref{eq:kdeada}) is designed to do. This leaves us with a computational challenge in that an adaptive kernel density estimate must be evaluated directly at all evaluation grid coordinates using the kernel contribution from all observations in the dataset. Clearly, this will be problematic for very large datasets and/or fine evaluation grid resolutions. In recent work, Davies \& Baddeley \cite{davbad:2017} examine possible solutions to this matter in some detail.

The first solution is rather simple: Approximate the adaptive kernel estimate $\hat{f}_{h_0}$ by a sum of fixed-bandwidth estimates operating on appropriate subsets of the observed data, defined by binning the variable bandwidths themselves. Specifically, for a given adaptive kernel density estimate of the spatial data set $\bX=\{\bx_1,\ldots,\bx_n\}$, let $\hat{\bh}=\{\hat{h}_1,\ldots,\hat{h}_n\}$ represent the bandwidths calculated via (\ref{eq:hu}) at each observation i.e.\ $\hat{h}_i\equiv h(\bx_i;f)$; $i=1\ldots n$. Then, let $0<\delta\ll 0.5$ be a ``quantile-step'' value such that $D=\delta^{-1}$ is an integer, thereby defining $D$ \emph{bandwidth bins} as
\[
\Big[\hat{h}^{(0)},\hat{h}^{(\delta)}\Big],\quad\Big(\hat{h}^{(\delta)},\hat{h}^{(2\delta)}\Big],\quad\ldots,\quad\Big(\hat{h}^{(\{D-1\}\delta)},\hat{h}^{(1)}\Big],
\]
where $h^{(p)}$ denotes the $p$th empirical quantile of $\hat\bh$. The partitioning technique computes
\begin{equation}\label{eq:partit}
\hat{f}_{h_0}(\bx|\bX)\approx n^{-1}\left\{n_{(1)}\tilde{f}_{\bar{h}_1}(\bx|\bX_1)+\ldots+n_{(D)}\tilde{f}_{\bar{h}_D}(\bx|\bX_D)\right\};\qquad\bx\in W,
\end{equation}
where $\bX_d$ represents the $n_{(d)}$ observations---the subset of the data $\bX$---that are allocated to the $d$th bandwidth bin according to the variable bandwidths attached to each point, and $\bar{h}_d$ is the midpoint of the corresponding bin. With each component of the above sum being a fixed bandwidth estimate, the DFT approach may be applied $D$ times to yield the approximation to the final adaptive estimate. Numerical experiments in \cite{davbad:2017} showed (\ref{eq:partit}) to perform well in terms of accuracy to the direct, edge-corrected estimate $\hat{f}_{h_0}$, even with relatively coarse binning (e.g. $\delta=0.05$ for $D=20$ bins), with a noticeable reduction to computational expense.

By default, \verb|bivariate.density| in \verb|sparr| computes adaptive density estimates directly. However, if speed is necessary, the user can make use of the optional \verb|davies.baddeley| argument for a partitioned estimate. The easiest way to do so is to supply a single value to the argument, which is taken to represent $\delta$ in the notation above. We note values of $0.01$ (100 bandwidth bins) $0.025$ (40 bandwidth bins) and $0.05$ (20 bins) tend to provide good approximations, in decreasing order of execution time.

\begin{figure}[hbpt]
\centering
\includegraphics[width=0.3\textwidth,angle=-90]{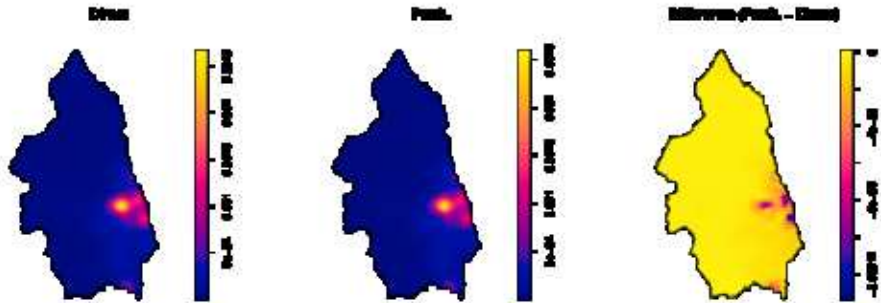}
\vspace{-10mm}
\caption{Comparing a direct (left) and a partitioned (middle) adaptive kernel density estimate of the pooled PBC data; the difference between the two given on the right.}\label{fig:partit}
\end{figure}

With \verb|sparr| and the PBC data loaded (the latter with a call to \verb|data(pbc)|), consider the next two lines, each of which calculates an adaptive estimate:

\small
\begin{verbatim}
R> system.time(direct <- bivariate.density(pbc,h0=4,hp=3,adapt=TRUE,verbose=FALSE))
   user  system elapsed 
  4.191   0.804   5.004 
R> system.time(partit <- bivariate.density(pbc,h0=4,hp=3,adapt=TRUE,davies.baddeley=0.025,verbose=FALSE))
   user  system elapsed 
  0.522   0.079   0.603 
\end{verbatim}
\normalsize
As such, for an adaptive density estimate of the pooled PBC data, setting $\delta=0.025$ took a little over half a second to complete, with the elapsed time for a direct estimate extending to around 5 seconds on the lead author's desktop machine. The three plots in Figure \ref{fig:partit} show both the direct and partitioned estimates, and a pixel image of the difference between the two surfaces. The latter demonstrates relatively minimal overall discrepancy between the two versions.

While, as noted above, we cannot express the adaptive kernel estimator as a 2D convolution to allow Fourier transformation and fast evaluation, it was shown in \cite{davbad:2017} that by augmenting the dimension of our planar data $\bX$ to include the Abramson \cite{abram:1982} variable bandwidths (\ref{eq:hu}) themselves, we can construct a 3D kernel that does possess the required properties for a valid DFT. The basic idea is to log-transform the bandwidths $\hat{\bh}$ and combine them with the planar coordinates, forming a dataset in 3D space as $\bZ=\{(\bx_1,\log\hat{h}_1),\ldots,(\bx_n,\log\hat{h}_n)\}$. Then it can be shown, with the help of a special kernel function of the form
\begin{equation}\label{eq:K3d}
\mK(\bu,v)=\exp(-v)^{-2}K\left(\frac{\bu}{\exp(-v)}\right),
\end{equation}
that the convolution thereof with the point masses at $\bZ$ yields the adaptive kernel estimate at any valid range of global bandwidths \textit{simultaneously}. This is convenient, because it allows use of the 3D DFT, and once computed, it is possible to ``slice out'' any desired adaptive kernel estimate of the original spatial data $\bX$ virtually instantaneously. The fast-slice-lookup behavior makes the approach particularly useful for situations in which we wish to repeatedly evaluate an adaptive density estimate on the same dataset at different values of $h_0$, such as in bandwidth selection (for a full account of the technical details, see \cite{davbad:2017}).

Such \emph{multi-scale} adaptive estimation via the special kernel (\ref{eq:K3d}) is available in \verb|sparr| as the \verb|multiscale.density| function. With the same pilot and initial global bandwidth as used above, note a multi-scale density estimate of the pooled PBC data can be achieved with the following:

\small
\begin{verbatim}
R> pbc.multi <- multiscale.density(pbc,h0=4,hp=3)
Initialising...Done.
Discretising...Done.
Forming kernel...Done.
Taking FFT of kernel...Done.
Discretising point locations...Done.
FFT of point locations...Inverse FFT of smoothed point locations...Done.
[ Point convolution: maximum imaginary part= 8.63e-14 ]
FFT of window...Inverse FFT of smoothed window...Done.
[ Window convolution: maximum imaginary part= 8.44e-15 ]
Looking up edge correction weights...
1  2  3  4  5  6  7  8  9  10  11  12  13  14  
\end{verbatim}
\normalsize
By default, the function computes estimates over a discretised range of global bandwidths bounded by $0.25h_0$ to $1.5h_0$, where $h_0$ is specified in \verb|h0|. This range can be altered with the optional argument \verb|h0fac|. We can now recover any adaptive density estimate of the pooled PBC data we desire, corresponding to a $h_0$ within the range

\small
\begin{verbatim}
R> available.h0(pbc.multi)
[1] 1.041866 5.510191
\end{verbatim}
\normalsize
very quickly using the \verb|multiscale.slice| function. Each call below takes less than $\frac{1}{100}$th of a second, providing the results for estimates using $h_0=2.5$, $h_0=3.45$, and $h_0=5.3$ as \verb|bivden| objects, just as if we had computed them separately using \verb|bivariate.density|:

\small
\begin{verbatim}
R> den.a <- multiscale.slice(pbc.multi,h0=2.5)
R> den.b <- multiscale.slice(pbc.multi,h0=3.45)
R> den.c <- multiscale.slice(pbc.multi,h0=5.3)
\end{verbatim}
\normalsize
The user can either plot the above objects as usual, or execute \verb|plot(pbc.multi)| directly to view the discretised multi-scale density estimate, as an animation, from smallest $h_0$ to largest.

The extremely fast slice operation is very convenient for exploratory data analysis, and as noted above, for bandwidth selection methods. The implemented bandwidth selectors in \verb|sparr| for spatially adaptive kernel density estimates all rely on this computational advantage.

\subsection{Interactive 3D Plotting}\label{sec:3dplot}
Visualisation of spatial and spatiotemporal density and relative risk estimates is typically done with image plots as we have produced throughout this paper. Especially in the spatiotemporal setting, however, it can be useful to create 3D perspective plots of the function of interest, for a better visual impression of fluctuations and relative magnitudes over the study domain.

We can produce intricate 3D plots of bivariate and trivariate functions relatively easily with the aid of the contributed packages \verb|rgl| \cite{adletal:2017} and \verb|misc3d| \cite{fentie:2008}; both installed directly in \verb|R| via the command prompt as usual. These packages and their plotting functions are of note because they allow the user to interact with the finished product with the mouse; left-click and hold to rotate, right-click to zoom. 

\begin{figure}[hbpt]
\centering
\includegraphics[width=0.4\textwidth]{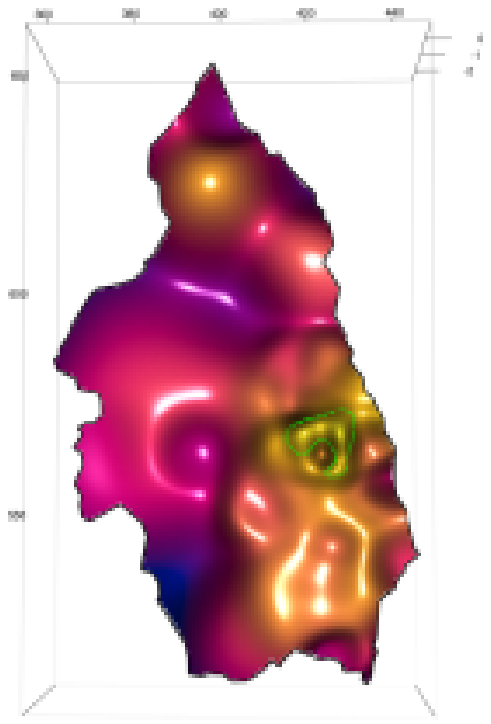}
\includegraphics[width=0.4\textwidth]{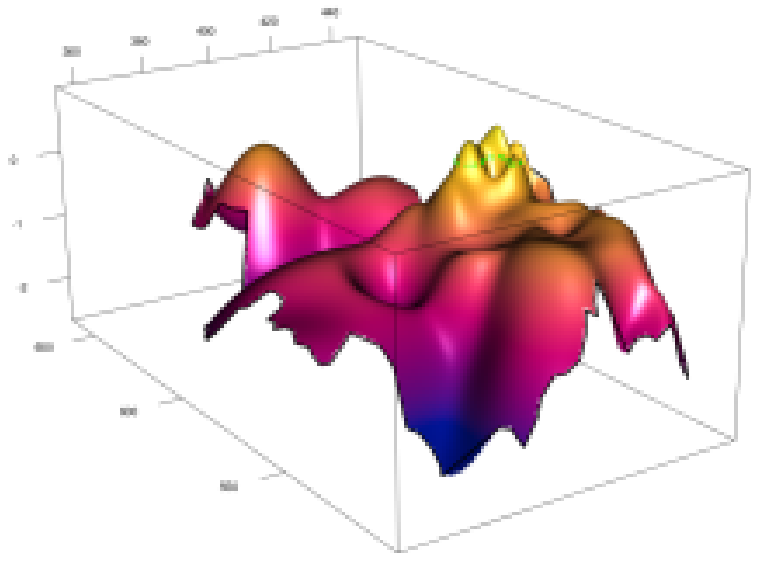}\\
\includegraphics[width=0.4\textwidth]{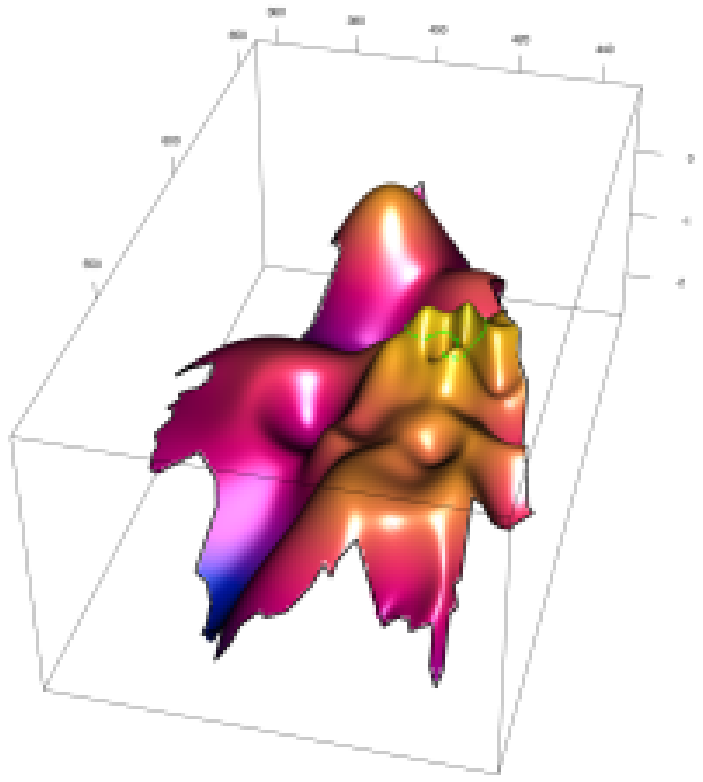}
\includegraphics[width=0.4\textwidth]{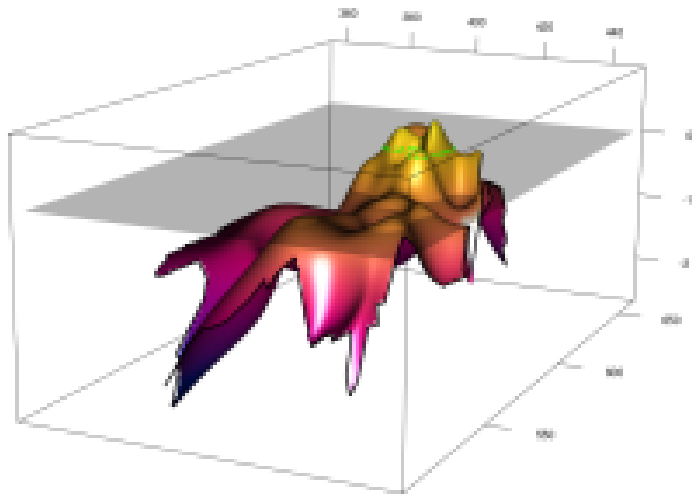}
\caption{Screenshots taken from different angles of an interactive 3D perspective plot of the symmetric adaptive spatial log-relative risk surface of the PBC data. Significantly elevated risk is highlighted by a green ASY contour at the 1\% level; and the bottom right screenshot shows an additionally included plane marking off the null log-risk at $0$.}\label{fig:pbc3d}
\end{figure}

Beginning with purely spatial relative risk, Figure \ref{fig:pbc3d} shows a collection of screenshots taken from a live, interactive plot of the symmetric adaptive kernel log-relative risk surface of the PBC data produced in Section \ref{sec:adacode}. The tolerance contour corresponding to the test for elevated risk at a significance level of $0.01$ is shown as a green outline. The bottom right screenshot shows an optionally added transparent plane to mark off the `null' risk at $\hat{\rho}_{h_0}=0$. We defer the code that produces this graphic to the supplementary material.

\begin{figure}[hbpt]
\centering
\includegraphics[width=0.49\textwidth]{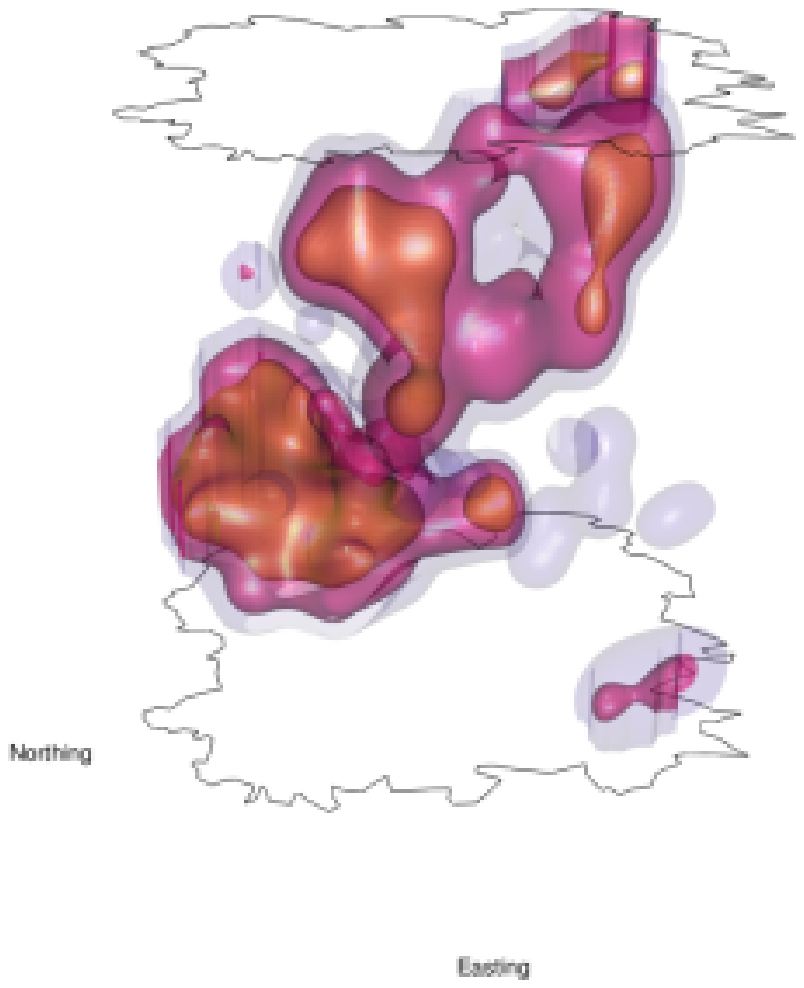}
\includegraphics[width=0.49\textwidth]{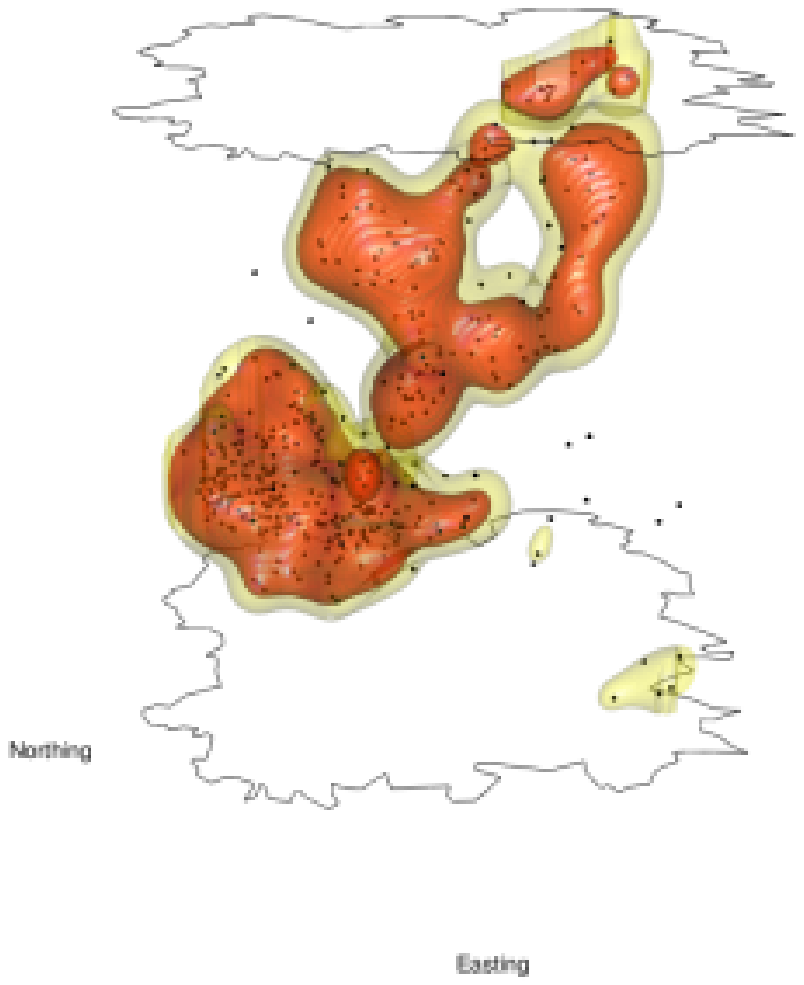}
\caption{Screenshots of interactive 3D plots depicting the estimated joint spatiotemporal log-relative risk of FMD during the 2001 outbreak in Cumbria. Left: The risk surface itself; isosurfaces drawn at log-risk levels of $0$, $1$ and $2$. Right: 3D contours delineating upper-tailed ASY tolerance contours at the 5\% (yellow) and 0.01\% (orange) levels of significance, along with the case observations.}\label{fig:fmd3d}
\end{figure}

Turning to the spatiotemporal domain, Figure \ref{fig:fmd3d} shows screenshots of a pair of similarly interactive graphics for the estimated joint spatiotemporal relative risk of FMD as estimated in Section \ref{sec:strrsparr}. The first plots 3D contours of the relative risk itself as isosurfaces at three different levels of log-risk, purple to pink to orange, in increasing opacity: $0$ (null raw risk $= 1$); $1$ (raw risk $\approx 2.72$); and $2$ (raw risk $\approx 7.39$). The second image shows a screenshot of the plot with isosurfaces delineating the tolerance contours for significantly elevated joint spatiotemporal log-risk at the weak and stringent significance levels of 5\% and 0.01\%; yellow and red respectively. The case observations are added to the latter graphic, and we once more defer \verb|R| code to the supplementary file. In both graphics, we observe the early bout of infections sparking heightened risk in the northwest, with the high risk moving further south as time increases.

\section{Conclusion and Ongoing Research Pursuits}\label{sec:conc}
Construction of continuous density-ratio or relative risk surfaces can be achieved in a very flexible way by employing kernel smoothing to estimate the requisite densities. It is most popular in geographical epidemiology, where it was originally developed as a tool to explore the spatial fluctuations in the risk of some disease after accounting for the heterogeneity of the underlying population, though has subsequently seen a steady rise in popularity across many different fields to address similarly posed research questions.

A number of novel methodological and computational developments followed the early work on the classical fixed bandwidth kernel estimator of relative risk, both to improve the estimates and the inferences we might draw therefrom, as well as to expand the domain of the data able to be considered. Of note is the use of spatially adaptive smoothing and the ability to compute spatiotemporal relative risk. 

Our ongoing research continues to target improvements to relevant statistical methods. Naturally, the familiar issue of optimal bandwidth selection remains an important consideration, and we are presented with a number of unique and difficult problems in this context. Further attention is warranted, for example, when it comes to optimal global and pilot bandwidth selection for spatially adaptive estimation of relative risk, a situation that is in part hindered by the additional computational expense of this approach. Similar comments apply to the estimation of spatiotemporal densities and relative risk functions---in particular, the best way to address optimal smoothing in the transdimensional time-static control estimator is presently unclear. Our concurrent research efforts are in part focussed on improving bootstrap-based bandwidth selectors, which we have noted in an \textit{ad hoc} sense to be more stable for these types of applications than leave-one-out approaches. Other relevant work includes testing for changes between two spatial relative risk surfaces (such as those obtained at different times); see \cite{haz:2017}.

In this tutorial we have reviewed the current state-of-the-science as it relates to kernel estimation of spatial and spatiotemporal density and relative risk functions, with a particular emphasis given to the practical aspects of implementation. This encapsulates a number of distinct techniques that demand careful consideration in a given analysis, such as fixed and adaptive smoothing regimens for spatial density estimation; inclusion of timestamps for spatiotemporal density estimation; edge-correction; relative risk and complications such as asymmetric versus symmetric estimators in the spatially adaptive setting and time-varying versus time-constant denominators in the spatiotemporal setting; bandwidth selection and jointly optimal bandwidth selection equipped to handle edge-correction (with further subtleties related to the type of density or risk estimator being employed); and the evaluation of Monte-Carlo and asymptotic $p$-value surfaces to highlight statistically significant fluctuations in an estimated risk surface. To improve accessibility to these specialised techniques for applied researchers, we have made the functionality discussed herein publicly available in the \verb|R| language via the new \verb|sparr| package, and we hope its release serves to both facilitate relevant pursuits by the research community as well as to open up further avenues for multidisciplinary research.

\section*{Acknowledgements}
The authors are grateful to the Animal and Plant Health Agency (APHA), UK, for permission to incorporate the anonymised FMD data into \verb|sparr|, and to P.J.\ Diggle for the PBC data. T.M.D.\ and M.L.H.\ acknowledge partial financial support by the Royal Society of New Zealand, Marsden Fast-start grant 15-UOO-092: \emph{Smoothing and inference for point process data with applications to epidemiology}.

\begin{appendices}

\section{Key \texttt{sparr} Content}\label{app:sparr}
\begin{center}
\footnotesize
\begin{tabular}{l||l|c|c}
\textbf{Function/Object} & \textbf{Description} & \textbf{Section(s)} & \textbf{Equation(s)}\\
\hline
\verb|bivariate.density| & Fixed and adaptive spatial kernel density estimation & \ref{sec:kdefix}, \ref{sec:kdeada}, \ref{sec:multi} & (\ref{eq:kdefix}), (\ref{eq:kdeada}), (\ref{eq:partit}) \\
\verb|BOOT.density|, \verb|BOOT.spattemp| & Bootstrap bandwidth selection for densities & \ref{sec:hboot} & (\ref{eq:boot}), (\ref{eq:bootada}) \\
\verb|fmd| & Anonymised FMD data & \ref{sec:mot}, \ref{sec:strrcode}, \ref{sec:3dplot} & -- \\
\verb|LIK.density|, \verb|LIK.spattemp| & Likelihood CV bandwidth selection for densities & \ref{sec:hlik} & (\ref{eq:lik}) \\
\verb|LSCV.density|, \verb|LSCV.spattemp| & Least-squares CV bandwidth selection for densities & \ref{sec:lscv} & (\ref{eq:lscv}) \\
\verb|LSCV.risk| & Jointly optimal bandwidths for spatial relative risk & \ref{sec:jointh} & (\ref{eq:joi1})-(\ref{eq:joi3}) \\
\verb|multiscale.density|, \verb|multiscale.slice| & Multi-scale adaptive spatial density estimation & \ref{sec:multi} & (\ref{eq:K3d}) \\
\verb|NS|, \verb|NS.spattemp| & Normal-scale bandwidth for densities & \ref{sec:hthumb} & (\ref{eq:hns}) \\
\verb|OS|, \verb|OS.spattemp| & Oversmoothing bandwidth for densities & \ref{sec:hthumb} & (\ref{eq:hosd}), (\ref{eq:hos}) \\
\verb|pbc| & PBC data & \ref{sec:mot}, \ref{sec:srrcode}, \ref{sec:3dplot} & -- \\
\verb|plot| & Plot various \verb|sparr| objects & \ref{sec:srrcode}, \ref{sec:strrcode} & -- \\
\verb|risk| & Spatial relative risk estimation & \ref{sec:ssrrat} & (\ref{eq:rhofix}), (\ref{eq:rhoada}) \\
\verb|spattemp.density| & Spatiotemporal density estimation & \ref{sec:st} & (\ref{eq:kdest}) \\
\verb|spattemp.risk| & Spatiotemporal relative risk and $p$-value surfaces & \ref{sec:tvden}, \ref{sec:tcden}, \ref{sec:sttol} & (\ref{eq:rrst})-(\ref{eq:rrstcond2}), (\ref{eq:hypstj})-(\ref{eq:varasyst}) \\
\verb|spattemp.slice| & Slicing a spatiotemporal density/relative risk estimate & \ref{sec:strr}, \ref{sec:strrsparr} & -- \\
\verb|tolerance|, \verb|tol.contour| & ASY and MC $p$-value surfaces for spatial relative risk & \ref{sec:tolmc}, \ref{sec:tolasy}, \ref{sec:srrcode} & (\ref{eq:hyp})-(\ref{eq:varadasym})
\end{tabular}
\normalsize
\end{center}

\end{appendices}

\bibliographystyle{plain}
\bibliography{literature}

\end{document}